\documentstyle[11pt,aaspp4]{article}
\def\etal{{\it et~al.\/\ }}
\def\etals{{\it et~al.'s\/\ }}

\def\teff{{\it T$_{\rm eff}$}}
\def\logg{{\rm log~g}}

\def\kms{{km~s$^{-1}$}}

\def\bd{{BD+56$^\circ$\,576}}

\slugcomment{In preparation for the Astrophysical Journal}
\lefthead{Venn et al.}
\righthead{Boron in B-type Stars}

\begin{document}

\title{Boron Abundances in Main Sequence B-type Stars: \\
       A Test of Rotational Depletion during Main Sequence Evolution
       \footnote{Based on observations made with the NASA/ESA 
        {\it Hubble Space Telescope}, obtained at the Space Telescope 
        Science Institute, which is operated by the Association of 
        Universities for Research in Astronomy, Inc., under NASA 
        contract NAS 5-26555. These observations are associated 
        with proposal GO\#07400.}}

\author{K.\,A.\,Venn\altaffilmark{2} and A.\,M.\,Brooks} 
\affil{Macalester College, Saint Paul, MN, 55105}

\author{David\,L.\,Lambert}
\affil{Univ.\,Texas at Austin, Austin, TX, 78712}

\author{M.\,Lemke}
\affil{Dr.\,Karl Remeis Sternwarte, Bamberg, Germany}

\author{N.\,Langer}
\affil{Utrecht, Netherlands}

\author{D.\,J.\,Lennon}
\affil{INT, La Palma}

\and 

\author{F.\,P.\,Keenan}
\affil{Queen's University - Belfast}

\altaffiltext{2}{University of Minnesota, 
Department of Astronomy, Minneapolis, MN, 55455}

\begin{abstract}
Boron abundances have been derived for seven main sequence B-type stars 
from {\it HST} STIS spectra around the \ion{B}{3} 2066\AA\ line. 
In two stars, boron appears to be undepleted with respect to the 
presumed initial abundance.   In one star, boron is detectable but
it is clearly depleted.   In the other four stars, boron is undetectable
implying depletions of 1 to 2 dex. 
Three of these four stars are nitrogen enriched, but 
the fourth shows no enrichment of nitrogen.   
Only rotationally induced mixing predicts that boron depletions 
are unaccompanied by nitrogen enrichments.   
The inferred rate of boron depletion from our observations 
is in good agreement with these predictions.    
Other boron-depleted nitrogen-normal stars are identified 
from the literature.   Also, several boron-depleted nitrogen-rich
stars are identified, and while all fall on the boron-nitrogen 
trend predicted by rotationally-induced mixing, a majority have nitrogen
enrichments that are not {\it uniquely} explained by rotation.

The spectra have also been used
to determine iron-group (Cr, Mn, Fe, and Ni) abundances.
The seven B-type stars have near solar iron-group abundances, 
as expected for young stars in the solar neighborhood.
We have also analysed the halo B-type star, PG\,0832+676. 
We find [Fe/H]~=~$-$0.88 $\pm$0.10, and the absence of 
the \ion{B}{3} line gives the upper limit [B/H] $<-$2.5. 
These and other published abundances are used to infer the 
star's evolutionary status as a post-AGB star.

\end{abstract}


\section{Introduction}

The light trace elements lithium, beryllium, and boron (LiBeB) 
are at the center of
astrophysical puzzles involving sites as diverse as the primordial 
fireball, interstellar or even intergalactic space, and stellar 
surfaces and interiors.   This paper is concerned primarily with 
boron's role in testing models of stellar interiors and evolution.
This role arises because boron nuclei are destroyed by warm protons, 
and thus even quite shallow mixing of the atmosphere with the interior 
reduces the surface abundance by bringing boron depleted material to the 
surface.  Lithium and beryllium are similarly affected. 
Thus, a determination that the surface abundance of a light element 
is less than the star's initial abundance is an observational
constraint for testing models of stellar interiors.

Lithium has served to test models of cool stars at essentially all phases 
of evolution, from the pre-main sequence through to the post-AGB stage,
with the atomic resonance doublet conveniently placed at 6707\,\AA.
Beryllium has seen similar but limited use through the \ion{Be}{2}
3130\,\AA\ resonance doublet.   
Boron has seen little use in testing mixing and other processes 
(e.g., mass loss) owing to the location of its resonance lines of 
\ion{B}{1}, \ion{B}{2}, and \ion{B}{3} in the ultraviolet.
This is unfortunate because boron alone is observable in hot stars. 

In the case of hot stars, boron abundances have been determined using 
the \ion{B}{2} 1362\,\AA\ resonance line by Boesgaard \& Heacox (1978) 
using {\it Copernicus} data,  Venn, Lambert, \& Lemke (1996) using {\it IUE} 
data, and   Cunha \etal (1997) from the GHRS on the {\it Hubble
Space Telescope}.    However, in stars of mid-B and earlier spectral types, 
boron is predominantly present as \ion{B}{3}, not \ion{B}{2}.
Proffitt \etal (1999) was the first to use the 
\ion{B}{3} 2066\,\AA\ line to determine boron abundances and 
$^{11}$B/$^{10}$B isotopic ratios in three B-type stars from very high 
S/N {\it HST} GHRS spectra.   Subsequently, Proffitt and Quigley (2001) 
mined the {\it IUE} archive for high-resolution spectra of B-type 
stars, and reported boron abundances or upper limits for 44 early B-type stars. 

A principal goal of most of these studies of hot stars was to establish 
the present-day boron abundance in order to improve our understanding of the
galactic chemical evolution of boron.   In this paper, our primary goal 
is different; we are searching for boron-depleted stars to understand how 
their depletion arose.   Only Proffitt \& Quigley (2001) have shared 
this goal.  Herein, we shall describe a test of rotationally-induced mixing 
in rapidly rotating massive stars.  
Shallow mixing results in the destruction of boron (lithium and beryllium, also) 
by proton capture.  Deeper mixing also penetrates regions where
H-burning via the CN-cycle has converted carbon into nitrogen.   
From any deep mixing, one may thus anticipate a drop in the surface
abundance of boron, followed by a continued decline in boron accompanied 
by a nitrogen enrichment and carbon depletion as the star ages on the main sequence.  

New stellar evolution models that include the effects of 
rotationally-induced mixing (Heger \& Langer 2000, Maeder \& Meynet 2000) 
predict that boron is depleted during main sequence evolution.
Fliegner, Langer, \& Venn (1996) first predicted a boron-to-nitrogen 
relationship in rotating stars and suggested it as an observational 
test of rotational mixing in hot stars. 
This test is unique since boron and nitrogen abundance variations, 
and the boron-to-nitrogen relationship predicted, cannot be explained
by simple initial abundance variations, nor binary evolution with mass 
transfer (e.g., Wellstein \etal 2001, Wellstein 2000, 
discussed further below).
Extraordinary mass loss rates for B-type main sequence stars are
demanded if surface boron depletions are to occur through exposure
of boron depleted layers.   The lifetime of $^{11}$B and $^{10}$B
against proton capture equals the main sequence lifetime of about 
10$^7$ years at an internal layer with a temperature of about 
7 x10$^6$~K.   Above this layer resides about $\sim$1~M$_\odot$ 
of material (see Fig.~1 in Fliegner, Langer, \& Venn 1996,
and Fig.s~2, 3, \& 4 in Heger \& Langer 2000), and therefore
a mass loss rate of $\sim$10$^{-7}$ M$_\odot$/yr is required if
surface boron depletions are to occur.   Such a rate is typical
of B-type supergiants, but at least an order of magnitude larger
than derived for main sequence OB stars (Cassinelli \etal 1994).
The new rotating stellar evolution models also address other long-standing 
problems (beyond abundance anomalies), such as the origins of B[e] and 
WNL/Ofpe (slash) stars, the distribution of blue-to-red supergiants in 
the HR diagram, and the unseen post-main sequence gap predicted in all
standard stellar evolution scenarios (see Langer \& Heger 1999).

In this paper, we present boron abundances from the \ion{B}{3} 2066\,\AA\
line from new {\it HST} STIS spectra of seven Galactic main sequence B-type 
stars.   
These stars were selected from a variety of OB associations, and have been 
well-studied in the optical so that atmospheric parameters and surface
nitrogen and carbon abundances are available in the literature.   
We have also used the spectra to determine their iron-group abundances, 
which are difficult to determine accurately from optical studies of B-type
stars,
due to a lack of \ion{Fe}{3} lines and uncertainties in NLTE effects on
available lines.

Our observing program included an eighth star:  PG\,0832+676, a B-type star
in the halo, whose evolutionary status has been debated for a number of years. 
Brown \etal (1989) deduced that the star was a normal Population~I B1~V 
object at a galactocentric distance of R$_G \sim$ 37\,kpc, 
and $z~\sim$ 18\,kpc from the galactic plane.  Hambly \etal (1996)
reanalysed this star and found that $\alpha$-elements (Mg, Al, Si, S) 
were underabundant by about 0.4 dex relative to HR~1886, a local B-type 
star.
These authors argued that the star was not a H-burning star, but rather
was either a blue horizontal branch or a post-AGB star.
Our spectra provide not only an estimate of the boron abundance but also
new iron-group abundances (Hambly \etal detected one \ion{Fe}{3} line,
which suggested [Fe/H]$\sim-$0.5 from a differential analysis).  We
will use these abundances to further establish the star's evolutionary 
status.   Also, as Hambly \etal noted, PG\,0832+676 is extremely sharp-lined, 
and as such its ultraviolet spectrum could provide a superb template for line 
identifications and wavelength calibrations.

\section{Target Selection and Observations}

Eight B-type stars were selected for {\it HST} STIS spectroscopy near 
the \ion{B}{3} 2066\AA\ line, see Table~\ref{obs}.   
Five of these targets are bright stars in OB associations.
Our goal was to select well-studied stars with a range of 
carbon and nitrogen abundances, and located in OB associations 
of different ages, in order to search for trends in the boron 
abundances with these parameters.  
A sixth target is the bright star, HD\,34078, a
runaway OB star from Ori~OB1. 
These six bright objects are also known variable stars 
(typically, $\beta$ Cep variables - this seems to have no 
influence on the surface abundances, discussed below), 
and two are in known binary systems, see Table~\ref{varbin}. 
Two fainter targets are also included;
\bd\ in Per OB1, and PG~0832+676 in the Galactic halo.
All of our targets have well determined atmospheric parameters, 
see Table~\ref{atms}.   

Careful planning was necessary to avoid the MAMA-NUV brightness 
limits in observing the six bright targets.
Even the most narrow slit, with the high resolution E230H grating
(R~=~114,000), was rarely permitted for our observations.   This
combination worked only for HD\,34078 at V~=~6.0.   For the other 
brighter objects, it was necessary to use either the lower resolution 
E230M grating (R~=~30,000), or E230H with a neutral density filter, 
0.2x0.05ND (ND=2), to obtain our spectra in the minimum number of 
orbits; see Table~\ref{obs}.    The E230M grating is sufficient to 
resolve the \ion{B}{3} line in all of our targets; the instrumental 
broadening for this grating is 9~\kms\ ($\pm$1~\kms\ depending on 
the slit choice), whereas the targets have measured rotational
velocities ranging from 1$\le v$sin$i \le$39~\kms.
For the faint star, PG\,0832+676, the E230M grating and 0.2x0.2
slit combination produced a good quality spectrum for the abundance
analysis, but we were not able to take full advantage of the 
sharp-lined nature of PG\,0832+676 as a UV spectral template.

\begin{figure}
\plotone{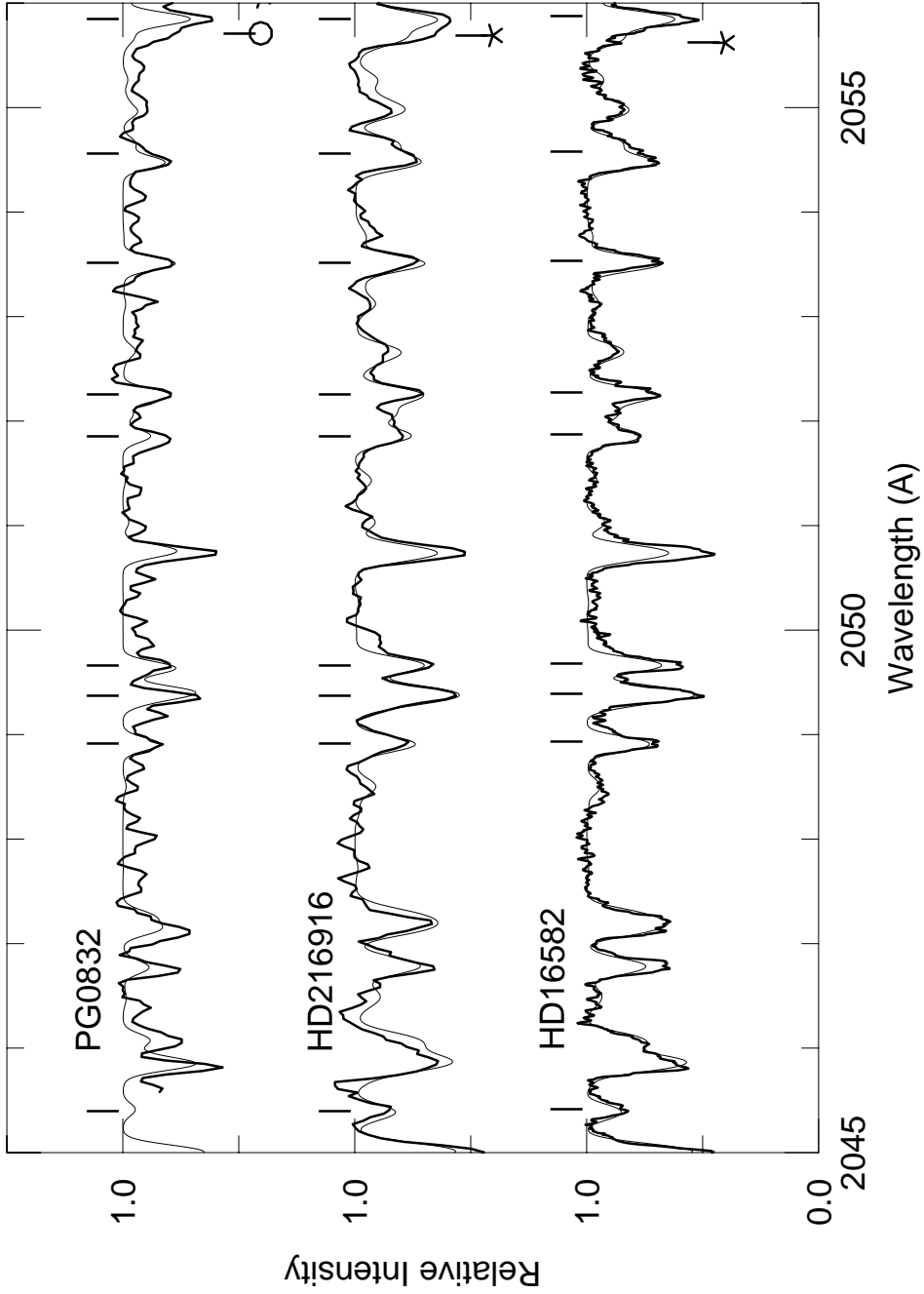}
\caption{Coadded {\it HST} STIS spectra for three B-type stars ({\it thick line})
and their spectrum syntheses ({\it thin line}).   
The observed spectra were smoothed for a 3 pixel resolution element, and
the higher resolution from the E230H grating can be seen for HD\,16582.
The iron-group metallicities in Table~\ref{metals} were used for each synthesis.
Features listed in Table~\ref{metals} are identified.  
Interstellar lines are marked; PG\,0832+676 has two sets of IS lines. 
\label{spec1}}
\end{figure}

\begin{figure}
\plotone{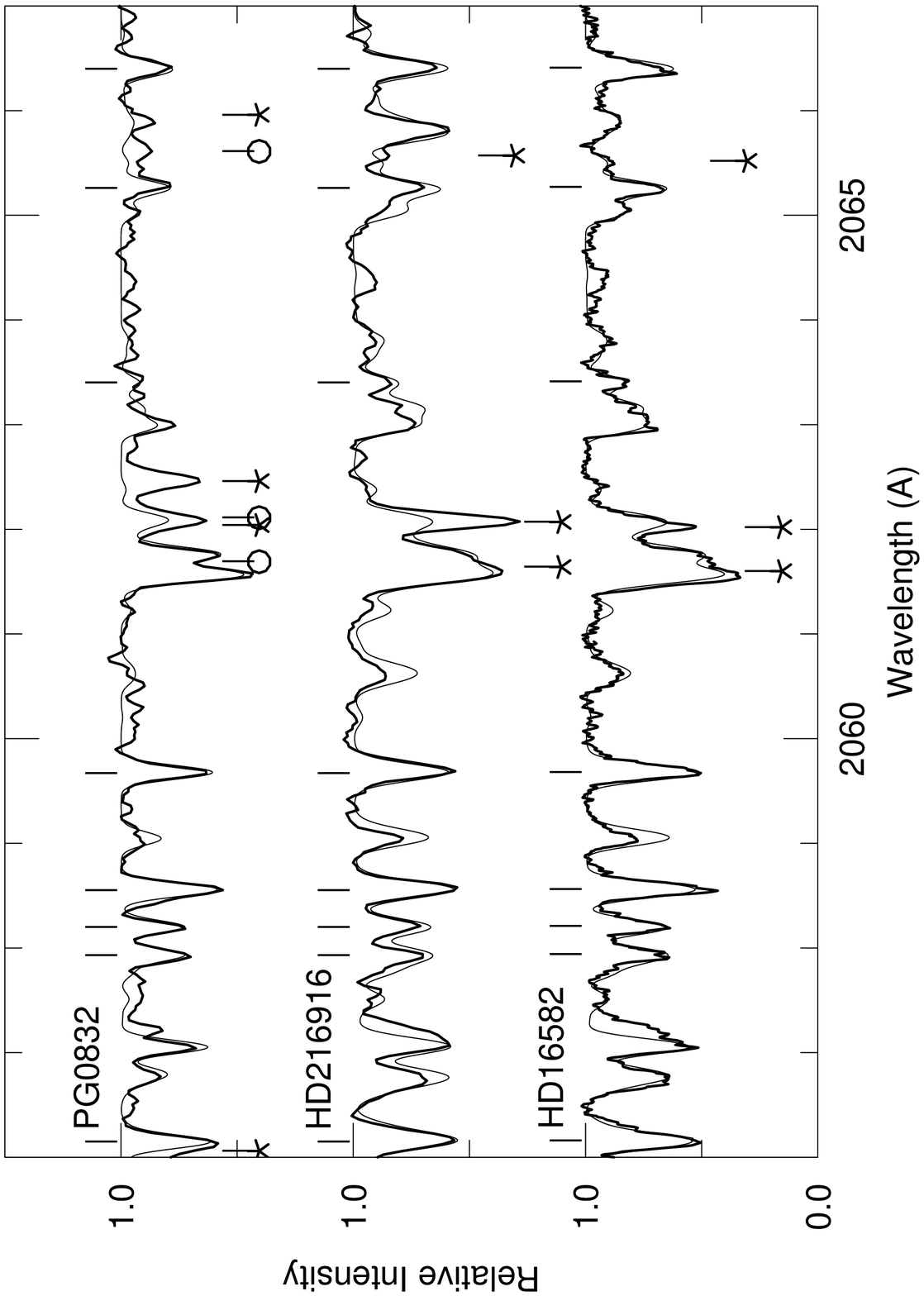}
\caption{See comments in Figure~\ref{spec1}.
\label{spec2}}
\end{figure}

\begin{figure}
\plotone{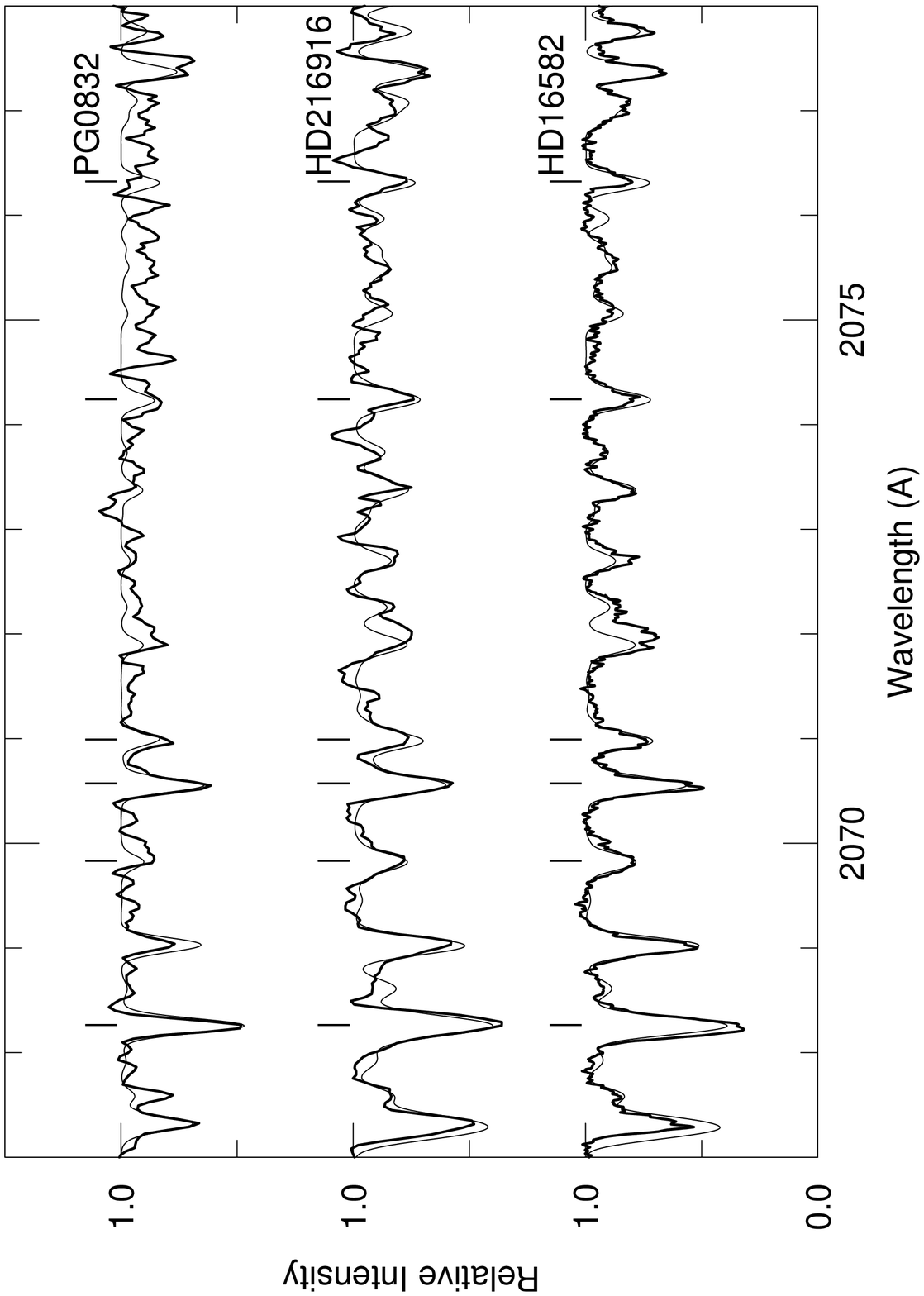}
\caption{See comments in Figure~\ref{spec1}.
\label{spec3}}
\end{figure}

Multiple exposures were taken at slightly varying central wavelengths
to maximize the S/N at the \ion{B}{3} 2066\AA\ line for most objects.
The exposure times and central wavelengths are listed
in Table~\ref{obs}.   Spectra were reduced using the STIS pipeline. 
Each echelle observation included between 34 and 52 orders.
We selected the order(s) that included the \ion{B}{3} line in
each exposure, and combined these for maximum S/N near 2066~\AA.
The spectra were smoothed (3-pixel boxcar smoothing), rectified 
using a low-order Legendre polynomial, and offset from vacuum
wavelengths (observed) to air wavelengths (line list, discussed below).
Sample spectra are shown in Figures~\ref{spec1}, \ref{spec2}, and \ref{spec3}.   
The resultant peak S/N values near 2066~\AA\ are listed in 
Table~\ref{obs}.   The final spectra used for iron-group and boron 
abundance determinations in this paper range from 2045 to 2078~\AA, 
with a dispersion ranging from $\sim$4 (E230H) to $\sim$16~\kms (E230M).

Four of the stars in this study are known $\beta$ Cep variable stars 
(see Table~\ref{varbin}).  In fact, HD\,205021 {\it is} $\beta$ Cep.  
These stars exhibit variability in 
their radial velocities as well as their luminosities.  For 
instance, HD\,205021 displays a peak-to-trough amplitude of 
$\sim$40~\kms over a period of 0.19~d (Hadrava \& Harmanec 1996).  
This period and amplitude is typical of the $\beta$~Ceph 
variables in our target list, and is long enough to display 
radial velocity variations over the full range of our observations.
However, the observations were divided into very short exposures 
to avoid the MAMA brightness limits.   Thus, there is little 
significant broadening due to the radial velocity variations 
per exposure (typically about 4~\kms, although the longest 
exposure for HD\,16582 may be broadened by as much as 9~\kms). 
As an additional test, we examined the FWHM of our individual 
exposures and found insignificant differences in most cases.
In the case of HD\,216916, we were surprised to find no significant
differences in the FWHM when comparing the 262\,s, 880\,s, and the 
1200\,s exposures (although the amount of broadening due to radial 
velocity variations will also depend on the phase when the observation 
is taken).  
Finally, we shifted each spectrum into the stellar rest frame
before coaddition to minimize smearing effects.

\section{The Abundance Analyses}

Elemental abundances have been determined from LTE spectral syntheses 
and ATLAS9 model atmospheres (Kurucz 1979, 1988).  
Solar metallicity models were used throughout, with the exception 
of PG\,0832+676 for which [Fe/H]~=~$-$1.0 models were adopted.  
LTE spectral syntheses were made using the program 
LINFOR\footnote{LINFOR was originally developed by H.\,Holweger,
W.\,Steffen, and W.\,Steenbock at Kiel University.   It has
been upgraded and maintained by M.\,Lemke, with additional
modifications by N.\,Przybilla.}.

The stellar \teff, gravity, and projected rotational velocities 
($v$sin$i$) values were adopted from the literature, see Table~\ref{atms}.  
We note here that there is some uncertainty in the temperature scale
for B-type main sequence stars.   This is discussed further in Section~4.
Other parameters were determined from the syntheses, i.e., macroturbulence,
microturbulence ($\xi$), and radial velocity, see Table~\ref{abus}. 
Macroturbulence ($\xi_{Ma}$) was initially set to the instrumental 
broadening values, then increased to best fit the spectral lines.
In most cases, the macroturbulence is only 3 to 5~\kms\ larger than
the smoothed instrumental profiles, however HD\,205021 and HD\,16582 
required macroturbulence values of 8~\kms\ larger. Observations for
these latter two stars were made with E230H, which has a very sharp
instrumental profile, thus much of the broadening in these two
stars is likely due to their $\beta$~Ceph radial velocity variations
(discussed above).  Another star, HD\,34078, observed with E230H has 
a much lower macroturbulence and is not a known $\beta$~Ceph variable.   
A gaussian line-spread function was adopted throughout our spectrum
syntheses; while acceptable, we recognize that line spread functions 
for the echelle gratings are available from STScI, but we neglected 
this specific instrumental broadening component in our syntheses.   
 
$\xi$ was determined for each star using the \ion{Fe}{3} lines, 
requiring that strong and weak lines yield the same iron abundance.
Our $\xi$ values are significantly lower than those in the literature 
from analyses of optical spectral lines (i.e., O~II).   This is not uncommon, 
as previously found by Venn \etal (1996) and Cunha \etal (1997), where $\xi$ 
from UV analyses is significantly smaller than from optical analyses 
of the same star.    This probably reflects differences in the structure of 
the upper atmosphere where the UV spectrum forms.  Regardless, the value of 
$\xi$ has little impact on the derived \ion{B}{3} 2065.8~\AA\ line 
abundance (see below).

Only the \bd\ $v$sin$i$ value was determined from syntheses since the 
upper-limit in the literature was clearly too large.   \bd\ and
PG\,0832+676 were observed with the same instrumental set-up, thus 
the broadening parameters from the PG\,0832+676 analysis 
were adopted for the \bd\ analysis.  PG\,0832+676 is a known 
sharp-lined object ($v$sin$i$=1~\kms), thus any remaining broadening 
required for \bd\ was attributed to its projected rotation rate.

\subsection{Line List}

Our spectral line list and atomic data originated from the Kurucz 
(1988; CD-18) line list, including all lines in the the 
iron-group, light elements, and heavy elements lists, up to barium, 
and through the fifth ionization state from titanium onwards.  
We updated this line list by including the new wavelengths for eight 
\ion{Fe}{3} lines reported by Proffitt \etal (1999) from FTS laboratory 
measurements.   We also updated the atomic data from Kurucz's semi-empirical
values for 64 \ion{Fe}{3} lines from Ekberg (1993).    
Ekberg determined new wavelengths for 
the three lowest configurations of \ion{Fe}{3} from laboratory measurements 
using a sliding spark discharge and spectrograph, with some energy levels 
computed assuming spin-orbit coupling, and transition probabilities 
calculated for all combinations.

\begin{figure}
\plotone{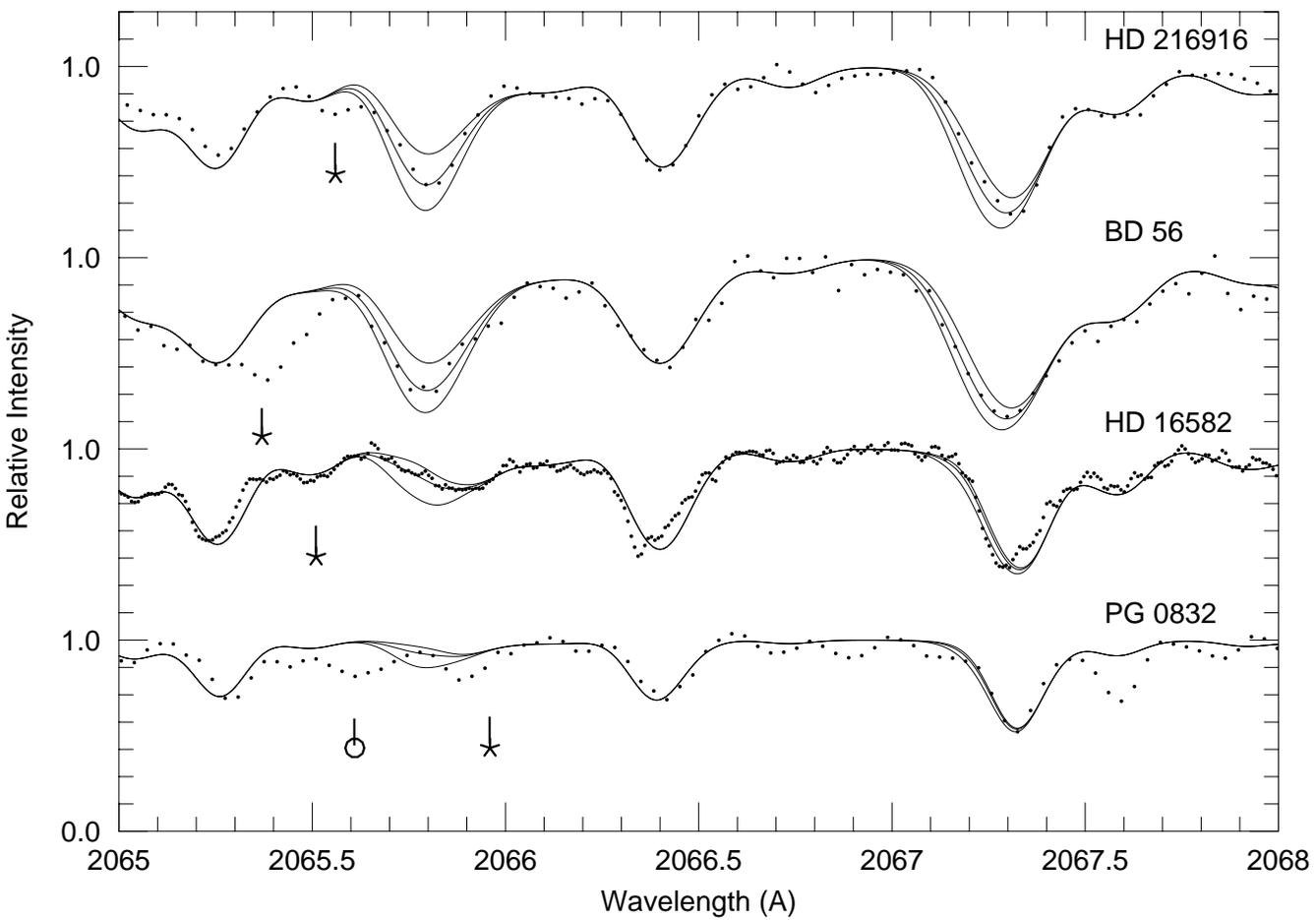}
\caption{Boron syntheses for the three stars where the \ion{B}{3}
feature is detected, and PG\,0832+676.   The best fit syntheses are
shown, as well as $\Delta$log(B/H)=$\pm$0.4 for comparisons.   
Interstellar lines are marked; PG\,0832+676 has two sets of IS lines. 
\label{bfita}}
\end{figure}

\begin{figure}
\plotone{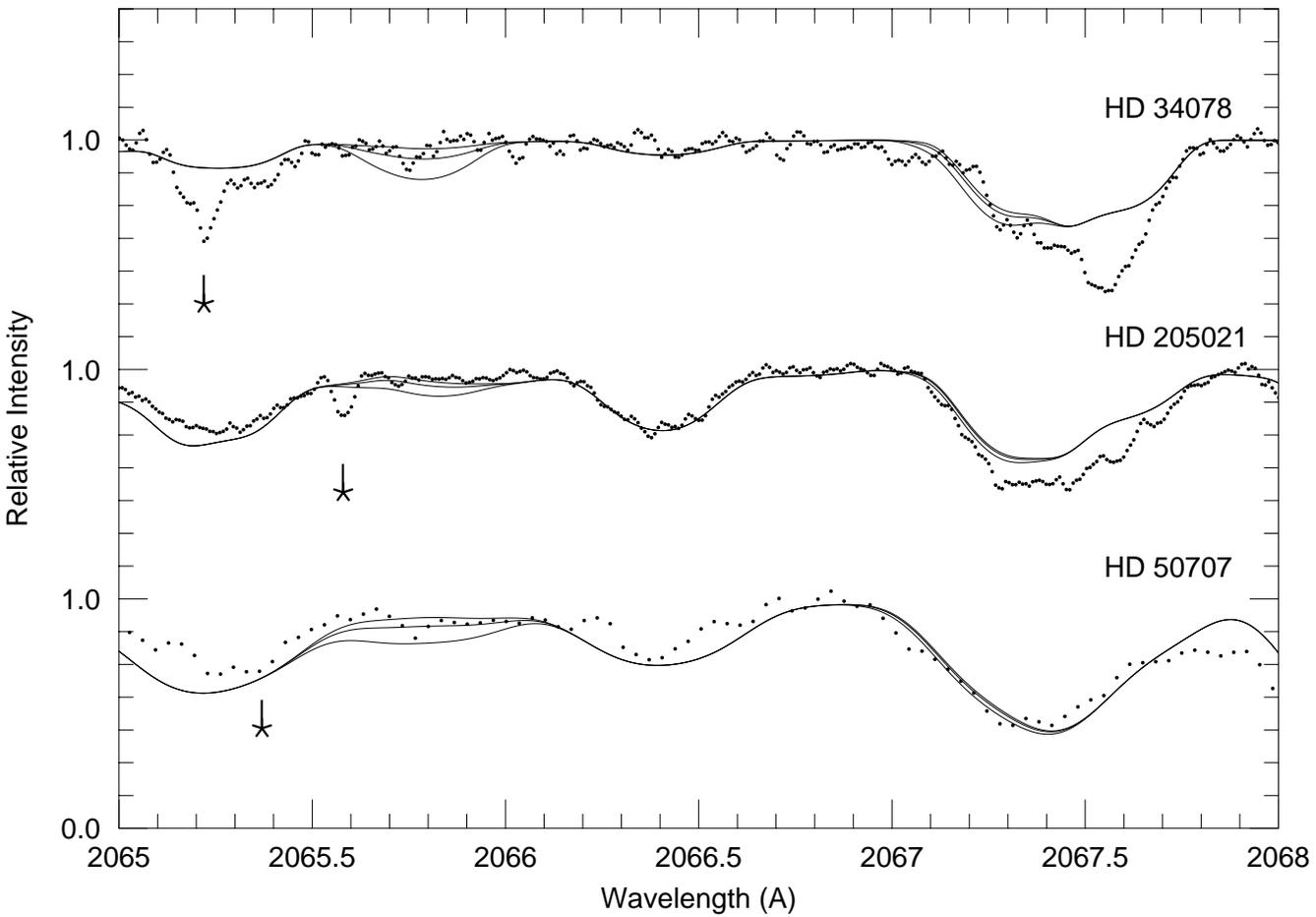}
\caption{Boron syntheses for three stars where the \ion{B}{3}
feature is {\it not} detected (see Figure~\ref{b365} for HD\,36591
synthesis).  The best fit syntheses are shown, as well as 
$\Delta$log(B/H)=$\pm$0.4 for comparison.   Clearly the boron
abundance is more difficult to determine and less accurate in
the broad lined stars.  Interstellar lines are marked. 
\label{bfitb}}
\end{figure}

\begin{figure}
\plotone{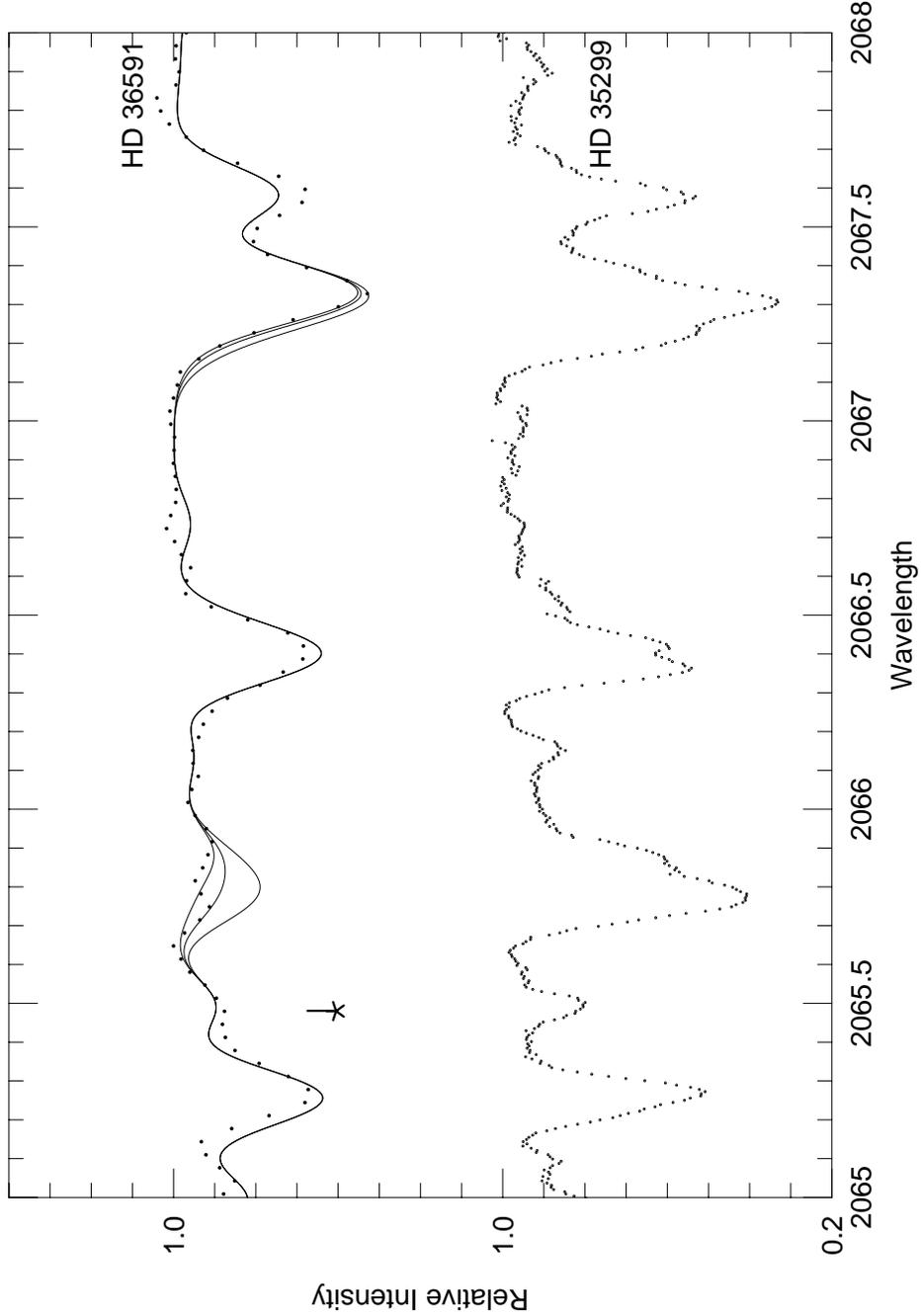}
\caption{Boron syntheses for HD\,36591; 
12+log(B/H)$_{\rm LTE}$=1.05, 1.45, 1.85 are shown.
Boron is clearly weak in HD\,36591.  
For comparison, an {\it HST} GHRS spectrum for 
HD\,35299 is shown.  These two stars are both in the
Orion OB1 association and have very similar atmospheric 
parameters, yet Proffitt \etal (1999) and Lemke \etal (2000) 
fit a normal LTE boron abundance to HD\,35299 
(12+log(B/H)=2.5 and 2.7, respectively).
Interstellar lines are marked for HD\,36591. 
\label{b365}}
\end{figure}

Data for the \ion{B}{3} 2${s^2}$S $-$ 2${p^2}$P resonance doublet 
with lines at 2065.8\,\AA\ and 2067.3\,\AA, and hyperfine
and isotopic components, are taken from Proffitt \etal (1999). 
The 2065.8\,\AA, the stronger and less blended of the doublet, 
is our primary indicator
of the boron abundance.   This feature is blended with a weak \ion{Mn}{3} 
line about 0.12\,\AA\ from the $^{11}$B line and 0.08\,\AA\ from the 
$^{10}$B line. In our spectra, the \ion{B}{3} and the \ion{Mn}{3} lines 
are not resolved.  Proffitt \etals high S/N and high resolution 
{\it HST} GHRS spectra of HD\,35299 and HD\,886 clearly show the 
contributions of the $^{10}$B, $^{11}$B, and \ion{Mn}{3} lines separately.
Their synthesis results for these two stars suggest that the \ion{Mn}{3} 
wavelength should be offset by 6\,m\AA\ redward of the Kurucz wavelength; 
also, they found that the abundance [\ion{Mn}{3}]~=~$-$0.2 from that one
feature, suggesting the gf-value is 0.2~dex too large.   
We adopt the \ion{B}{3} line list and transition probabilities 
given by Proffitt \etal (1999), and adjust the wavelength and
oscillator strength of the blended \ion{Mn}{3} line. 
[The weaker \ion{B}{3} line at 2067.3\,\AA\ is blended with a strong
\ion{Fe}{3} line and weak \ion{Mn}{3} line. The blending is of such 
severity that the 2067.3\,\AA\ stellar line is not suitable for a boron 
abundance determination, although we note that the best boron fits 
to HD\,216916, \bd, and the upper-limit for HD\,36591 produce the best 
fit to the 2067.3\,\AA\ B-Fe-Mn blend as well 
(see Fig.~\ref{bfita}, \ref{bfitb}, and \ref{b365})]. 

For all syntheses, we  assume an isotopic ratio $^{11}$B/$^{10}$B~=~4.0, 
the solar system ratio (Shima 1963), consistent with the
estimates given by Proffitt \etal (1999) from their line profile analyses
of two sharp-lined B-type stars.   We also consider a smaller ratio
in our boron uncertainty estimates (discussed below).  

Proffitt \& Quigley (2001) first noted the importance of the IS lines in this
wavelength range, particularly one \ion{Cr}{2} IS line that can come close to 
the \ion{B}{3} $\lambda$2065.8 feature depending on the stellar radial 
velocity.    Since interstellar lines are quite sharp, we have identified 
their rest wavelengths from our spectrum of HD\,205021.  This star has a 
high S/N and rotation rate that make the broad stellar lines quite 
distinct from the sharp IS lines
(e.g., see Fig.~\ref{bfitb}).   We have identified four IS lines
as \ion{Cr}{2} (2055.60, 2061.58, and 2065.50) and \ion{Zn}{2} (2062.01)
from Morton (1991). 
We note the locations of the IS lines in our spectrum figures
(see Fig.s~\ref{spec1} to \ref{b365}), where the IS
lines appear at different wavelengths because the (stellar) rest
frame is used.   
The IS line near the \ion{B}{3} feature is only a problem in PG\,0832+676.
PG\,0832+676 has two sets of IS lines, one of which appears to be blended with
the \ion{Mn}{3}-\ion{B}{3} 2066\,\AA\ feature.
This affects the accuracy of our upper limit for boron in PG\,0832+676, 
but nevertheless, the boron feature is clearly absent (see Fig.~\ref{bfita}).
We also note in this figure that a \ion{Ti}{4} feature near 2067.6\,\AA\
in PG~0832+676 is not well fit, but this is also seen in the spectrum of
the hot star HD\,34078, suggestive of atomic data uncertainties. 

Our final line list includes 2275 features between 2045 and 2078~\AA. 
All were included in the syntheses but many are negligible contributors.
A few final fine adjustments were made to the line list.   We examined the 
preliminary syntheses of three of the sharp-lined stars in this sample, 
HD\,216916, HD\,36591, and PG\,0832+676, and made slight wavelength shifts 
to improve the final spectrum syntheses.   These fine adjustments are 
reported in Table~\ref{offset-lines}.   We made no fine adjustments to
the oscillator strengths.
It is worth noting that our final line list does a remarkably good job 
at fitting the UV spectrum in our stars.  This is noteworthy because UV 
line lists are notoriously incomplete and/or uncertain in their atomic data.   
Examination of our spectrum figures show very few missing lines and 
quite good fits, suggesting that the energy levels and transition 
probabilities are fairly accurate.

\subsection{Iron-group \& Synthesis Parameters}

The iron-group abundances were determined from examination 
of a number of clean features in the spectra.  Spectrum syntheses 
of these individual features were performed, tabulated, and the 
results averaged. This allowed us to minimize
uncertainties due to poorly fit lines, as well as lines 
which appear to have poor atomic data.
In Table~\ref{metals}, we list the abundances relative to
the meteoritic abundances from Grevesse \& Sauval (1998), 
i.e., log(Fe)~=~7.50 and log(Mn)~=~5.53.
The mean abundances in Table~\ref{metals} were calculated
by excluding line abundances that fall more than 2$\sigma$ (from the 
line-to-line scatter) from the mean.

We estimate an uncertainty of approximately $\pm$2~\kms\ in
the macroturbulence, based on line profile fitting, 
but only $\pm$1~\kms\ in microturbulence based on line strengths.
Effects of these uncertainties on the iron-group abundances are 
quantified in Table~\ref{metal-unc} for a few representative stars.   
Uncertainties were found by fitting the iron-group features listed
in Table~\ref{metals} individually, then averaging the changes.
Continuum placement and the adopted $\xi$ and macroturbulence
parameters dominate the errors in the iron-group abundances.   
NLTE effects are neglected throughout this iron-group
analysis.  Iron-group abundances are determined primarily 
from lines of the dominant species of the elements, e.g., \ion{Fe}{3},
whose atomic statistical equilibrium rarely suffer from NLTE effects.
 
We find the mean iron-group abundance of [Fe/H]~=~$-$0.06 $\pm$0.15 
from 6-23 features in seven main sequence B-type stars.
This is similar to Cunha \& Lambert's (1994) results
from 1-8 optical lines of \ion{Fe}{3} in 16 Orion OB1 B-type stars,
[Fe/H]~=~$-$0.05 $\pm$0.10, and also to Gies \& Lambert's (1992) 
results from 1-3 optical lines of \ion{Fe}{2} or \ion{Fe}{3}   
in 31 Galactic B-type stars, [Fe/H]~=~+0.22 $\pm$0.20.
Six stars have {\it IUE} spectra that were examined by
Proffitt \& Quigley (2001) for an overall scaling factor 
for heavy element abundances.   Although they note that
the scaling factor per star should not be interpreted as a 
precise determination of the heavy element abundance, we do 
note that their values are only about 0.2\,dex lower than ours.

Finally, due to the uncertainties in the B-type main sequence
star temperature scale (discussed further in Section~4), we note
that a slight reduction in the Gies \& Lambert (1992) temperatures
(about 800~K, or 1~$\sigma$ in the \teff\ uncertainty) will reduce
our mean iron-group abundances by about 0.1~dex.

\subsection {Boron Abundances}

The LTE boron abundances listed in Table~\ref{abus} are
from spectrum syntheses, where we allowed the 
\ion{B}{3}~$\lambda$2065.8 and \ion{Mn}{3}~$\lambda$2065.9 
line abundances to vary independently in order to achieve 
the best possible fit to the observation.  
Spectrum synthesis fits for all our program stars 
are shown in Fig.s~\ref{bfita}, \ref{bfitb}, and \ref{b365}.
We did not attempt to constrain the \ion{Mn}{3}~$\lambda$2065.9 
line abundance, e.g., by setting the Mn abundance using other 
\ion{Mn}{3} lines, because of uncertainties in the atomic data 
for other \ion{Mn}{3} lines.    We report the
\ion{Mn}{3} 2065.9 line abundance determined from the best synthesis 
of the \ion{B}{3} blended feature in each star in Table~\ref{abus}.
It is in good agreement with the iron-group abundances per star.
 
To compute the uncertainty in the boron abundances, $\Delta$log(B/H),
independent of uncertainties in \ion{Mn}{3} 2065.9, we found it was 
necessary to fix the \ion{Mn}{3} 2065.9 line abundance {\it a priori}.   
Thus, we examined two methods (except for PG\,0832+676, 
discussed below).   Firstly, we simply set 
the value of \ion{Mn}{3} from the best fit synthesis 
(in Table~\ref{abus}), then allowed only the \ion{B}{3} 
line abundance to vary to calculate $\Delta$log(B).  
Secondly, we applied the mean iron-group corrections
(in Table~\ref{metal-unc}) 
to the \ion{Mn}{3} line component, and allowed boron to 
vary to find a new best fit for each uncertainty parameter. 
We found $\Delta$log(B) 
to be very nearly identical from these two methods for all
uncertainties investigated.  The largest differences between the
methods were $\le$0.04~dex for $\Delta\xi_{Ma}$ and
$\Delta\xi$.
In Table~\ref{boron-unc}, we report the uncertainties from the 
latter method.
Only for $\Delta$V$_{\rm rad}$ did we vary \ion{B}{3} 2065.8 and 
\ion{Mn}{3} 2065.9 together (both line abundance changes are
reported in Table~\ref{boron-unc}). 

For PG\,0832+676, we could not fit the \ion{B}{3}-\ion{Mn}{3} 
blend directly because of an interstellar line occuring 
near the location of the \ion{Mn}{3} line.   For this star, we
computed the iron-group abundances, and applied that result
for the \ion{Mn}{3} 2065.9 line abundance in order to determine 
a boron upper-limit.   We used this method since the 
\ion{Mn}{3} 2065.9 line abundance has been similar to the 
mean iron-group abundance in all of the other stars,
i.e., [Mn/Fe]~=~$-$0.05 to $-$0.35, see Table~\ref{abus} for
Mn and Table~\ref{metals} for Fe.   
If we applied a lower \ion{Mn}{3} 2065.9 line abundance
(e.g., [Mn/Fe]~=~$-$0.3), then the change in the boron 
{\it upper-limit} is small (e.g., $\Delta$log(B/H)~=~+0.1).
 
In summary, Table~\ref{boron-unc} shows that the 
most significant uncertainties in the boron abundances 
tend to be the continuum placement (thus, S/N of the data also)
and the radial velocity and assumed $^{11}$B/$^{10}$B ratio 
(discussed further below) may be significant in some cases.
Also, the \ion{Mn}{3} 2065.9\,\AA\ atomic data could play an
important role, but since the Mn abundance is not fixed for 
the boron syntheses, then uncertainties in the Mn atomic data 
will be compensated in the fit. 
Thus, the boron abundances should be accurate to $\pm$0.1 dex. 
It is worth noting that lowering the temperature scale (discussed
further in Section~4) has a negligible effect on the boron
abundances ($\Delta$log(B/H)~=~0.0 to $-$0.06).
 
Only for the hottest star, HD\,34078, is the derived boron abundance 
significantly sensitive to uncertainties in the atmospheric 
parameters.      In fact, examination of the predicted equivalent 
widths for the \ion{B}{3} 2065.8\,\AA\ lines in Fig.~\ref{beqw}
shows that the line strength is nearly constant between 20,000
to 28,000~K, then drops dramatically at higher and lower temperatures.  
For example, near 30,000~K, an uncertainty of $\pm$2000~K can change 
the boron equivalent width by $\mp$50\,\%.  From spectrum synthesis this 
corresponds to $\Delta$log(B/H)~=~$\mp$0.4, a substantial uncertainty
for only a 5~\% uncertainty in \teff.
For this reason, we examine our boron abundances 
versus temperature in Fig.~\ref{bt}; we also include boron
abundances from the literature on a homogeneous temperature
scale (discussed below).  
There is no significant trend, but we mark the temperature limits 
where the \ion{B}{3} 2065.8\,\AA\ line should be the most reliable.

Our LTE boron abundances in Table~\ref{abus} are corrected for NLTE 
effects.   For this paper, M.\,Lemke has extended his calculations 
(described in Cunha \etal 1997) to higher temperatures.  The NLTE
calculations were carried out for an abundance of 12\,+\,log(B/H)\,=\,2.6. 
As seen in Fig.~\ref{nlte}, the NLTE corrections for the \ion{B}{3} 
lines remain small, but reverse their sign in the hotter stars.  
This is because the ground state of \ion{B}{3} is overpopulated 
at the lower temperatures through collisional coupling and ionization
processes between \ion{B}{2} and \ion{B}{3}, and this is the only
mechanism that produces \ion{B}{3} NLTE effects at these temperatures.   
But as the temperature increases then the effects of \ion{B}{2} on the 
ground state of \ion{B}{3} are reduced.  At the same time, the UV radiation 
field increases and there is a general shift towards \ion{B}{3} as the 
main ionization stage such that photoionization dominates the \ion{B}{3} 
statistical equilibrium.   Eventually, this causes the sign reversal in 
the \ion{B}{3} NLTE corrections. 

Boron is clearly present and undepleted in only two stars 
in our sample: \bd\ and HD\,216916.    
That the boron abundance is quite different
in stars of very similar atmospheric parameters is
obvious from inspection of the spectra of HD\,216916 and HD\,16582.
For example, compare the large differences in the 
\ion{B}{3}-\ion{Mn}{3} 2066\AA\ blend in Fig.~\ref{bfita} 
for these two stars with insignificant differences 
in their other features (also see Fig.s~\ref{spec1}, \ref{spec2},
and \ref{spec3}).   Also, both of these stars are $\beta$ Cep
variables (see Table~\ref{varbin}), thus we conclude that the
$\beta$ Cep variability mechanism is not a cause of surface mixing.
Similarly, the Orion star HD\,36591 has similar atmospheric
parameters to another Orion star HD\,35299, yet vastly different 
boron line strengths are seen in Fig.~\ref{b365}.

Six of our stars have {\it IUE} spectra that were examined by 
Proffitt \& Quigley (2001).  Owing to the lower S/N and resolution 
of {\it IUE} spectra, their boron abundances are less precise than ours.
When boron is strong, as in HD\,216916, our results are in excellent
agreement and with similar uncertainties ($\pm$0.2 dex).    On the
other hand, Proffitt \& Quigley report detecting the boron line 
in HD\,34078, which we do not confirm (see Fig.~\ref{bfitb}).
For the other four stars in common, we report significantly 
lower abundances (HD\,16582) or lower upper-limits.

\section {Discussion}

\subsection{The Initial Abundance of Boron}

In order to pursue our principal goal of testing predictions
of boron-depletion due to rotationally induced mixing, it is necessary 
to establish the initial boron abundance of these local B-type stars. 
(PG\,0832+676 is discussed separately.)  
We suppose that the highest boron abundances
across our sample and others provide a close approximation to the
initial boron abundance.   We further assume that stellar boron abundances 
that are substantially less than the presumed initial abundance do 
not reflect an unusually low initial value, but rather arise from 
internal processes that deplete boron. 

The mean NLTE boron abundance from \bd\ and HD\,216916 is 
12+log(B/H)~=~2.3.   Proffitt \etal (1999)'s analysis of another 
two stars gives 12+log(B/H)~=~2.4 (considering NLTE corrections), 
and the mean NLTE abundance of stars with boron detections from 
the sample considered by Proffitt \& Quigley (2001) gives 
12+log(B/H)~=~2.4.
Cunha \etal (1997) found near solar boron abundances from
two B-type stars in the Orion OBIc subassociation, although
those results were from the \ion{B}{2} 1362~\AA\ line. 
Boron from the \ion{B}{1} 2497\,\AA\ line 
in F-G main sequence stars, with $-$0.15 $\geq$ [Fe/H] $\leq$ +0.15, 
is 12+log(B/H)~=~2.6 from Cunha \etal (2000a). 
 
These stellar abundances should be closely related to the
interstellar abundance of boron, as standard evolutionary 
models, as well as those including rotational effects (see below), 
predict survival of surface boron through the pre-main sequence phase.
Boron is detected in interstellar diffuse clouds by the 
\ion{B}{2} 1362\,\AA\ resonance line (Jura \etal 1996; 
Lambert \etal 1998; Howk, Sembach, \& Savage 2000).   
The lower limit to the interstellar abundance, 12+log(B/H)~$\geq$~2.4
suggested by Howk {\it et~al.}, is consistent with the above stellar 
abundances.  The lower limit is given in recognition of the possibility 
that boron may be depleted onto interstellar grains.   It is most
unlikely that boron is present {\it in} grains, if, as is widely
supposed, grains are formed primarily in stellar outflows from red
giants whose outer envelopes will be highly depleted in boron.

These stellar and nebular abundances are slightly lower 
than the solar system (meteoritic) 
value of 12+log(B/H)~=~2.78 (Zhai \& Shaw 1994).
For the B-type stars and the interstellar gas,
this reduction is in line with similar underabundances 
(relative to the Sun) for oxygen and other light elements. 
Since our primary interest lies in the B-type stars depleted 
in boron by an order of magnitude or more, the difference 
between an initial boron abundance of 2.4 and 2.8 is
largely irrelevant.

It is also worth investigating the boron abundance in the Orion
association as a fiducial point, since there are both stellar 
and interstellar determinations available.
In general, 2.4~$\le$~12+log(B/H)~$\le$~2.9 from B-type
stars and interstellar sight lines (among the references above). 
Some stars in Orion show large boron depletions though.
Two of our stars are or were members of the Orion association: 
HD\,36591 belongs to the Ib subgroup, and HD\,34078 (AE Aur) is 
a well known runaway star.   Neither star shows a detectable
\ion{B}{3} line.
While the boron depletion in HD\,34078 might be related to its 
transformation as a runaway star (e.g., contamination of the star 
by ejecta from its exploding companion), HD\,36591's low boron 
abundance is plausibly identified as arising from internal 
effects (discussed below).    This result is significantly
different from other boron determinations from B-type stars in
Orion (Cunha \etal 1997, Proffitt \etal 1999, Lemke \etal 2000).

Low boron abundances have been found in some other stars in Orion.
Cunha \etal (1999, 2000b) determined the boron abundances
in three G-type stars in Orion from the \ion{B}{1} 2497\,\AA\ line.
One of these, BD$-$5$^\circ$\,1317, resulted in an abundance 5x
less than solar, 12+log(B/H)~=~2.1 $\pm$0.2.   Surprisingly, this
star also proved to be oxygen rich (Cunha \etal 1998), 3x larger
than the Orion nebular abundance (Esteban \etal 1998). 
These results lead Cunha \etal (2000b) to propose a boron-oxygen 
anticorrelation, possibly the result of dilution of interstellar 
boron by relatively boron-poor but oxygen-rich ejecta from recent 
Type~II~SNe.   

Proffitt \& Quigley (2001 = PQ01) also discovered severely 
boron-depleted B-type stars, which muddles the 
issue of a boron-oxygen anticorrelation, particularly as 
the boron depletions are not related to oxygen enrichment.
In Fig.~\ref{boxy} (left panel), we show the boron-oxygen
anticorrelation from published data for Orion, 
as well as our star HD\,36591.
(We note that PQ01 include seven
more Orion stars with depleted boron, however five are hot 
stars like HD\,34078, which we consider less reliable boron
abundance indicators, and two have no published oxygen abundances).
In the right panel, we include only stars where boron is determined
from the reliable \ion{B}{3} feature 
(the boron and oxygen abundances plotted are from 
Table~\ref{bcno}, which is discussed further below).   
The existence of a boron-oxygen anticorrelation is not obvious 
from the \ion{B}{3} data alone, although we note that
much of the \ion{B}{3} 
data is from lower quality {\it IUE} spectra.  

The phenomenon of boron-depletion is not confined to the Orion 
association.    Proffitt \etal (1999) found that the field star 
HD\,3360 is boron-depleted, and we have found several boron-depleted 
stars in other OB associations, as did PQ01.    
Fliegner, Langer, \& Venn (1996)  first proposed that boron depletions 
may result from rotationally-induced mixing during the main sequence 
lifetime of B-type stars.   We discuss this in the next section.

\subsection{Rotationally-induced Mixing}

Recent models by Heger \etal (2000) follow the evolution of the angular 
momentum distribution and the occurence of associated mixing processes
in massive stars from the pre-main sequence through core collapse.    
They find that rotational mixing 
can affect stellar surface abundances, stellar life times, and the 
evolution of a star across the HR diagram (Heger \& Langer 2000 = HL00).
These new models have fundamental consequences for the age determination 
of young stellar clusters, and they can reconcile the long-standing 
mass discrepancy problem (that spectroscopic masses differ from those 
inferred from stellar evolution tracks, Herrero \etal 1992). 

In general, a rotating star has a lower effective gravity, thus it acts 
like it has less mass at core-H ignition.    
Later, during core-H burning, rotationally induced mixing of protons
from the envelope into the convective core and of helium from the
core into the envelope will lead to higher luminosities compared to
non-rotating models.
Also, the effective temperature is lower, thus the ZAMS position 
on the HRD changes from a non-rotating model.
Evolution on the main sequence then depends on the star's mass and
rotation rate, as well as the efficiency of mixing in the upper convective
core and stellar interior.   
 
Heger \& Langer's models cover a mass range of 8 to 25 M$_\odot$, 
and a range of rotational velocities from 0 to $\sim$450 \kms.  
They computed two sets of models, with different assumptions on
the efficiency of rotational mixing in layers containing a gradient
in the mean molecular weight $\mu$ (called a $\mu$-barrier).
In one set, the $\mu$-barriers are ignored, and the overall
efficiency of rotational mixing had to be reduced in order to
meet observational constraints.   In the other set, $\mu$-barriers
are taken into account, and again the overall mixing efficiency
above the $\mu$-barrier is set by observational constraints
(see Heger \etal 2001 for details).
The differences between the two sets of models reflects the
remaining uncertainties in the theoretical description of the
rotational mixing processes. 
Finally, it is interesting to note that even though the rotationally 
induced $\mu$-barrier {\it inhibits} mixing just above the core, the 
current models show that the envelope above the $\mu$-barrier is very
well mixed, thus CN-processed gas is {\it more} evident up through the 
photosphere then when the $\mu$-barrier is ignored.
 
The range in velocities examined by the models is thought to be 
typical for B-type stars. For example, Fukuda's (1982) statistical 
study of rotational velocities found B2~V and B2~IV stars with mean 
$v$sin$i$ values of 154 and 110~\kms, respectively, and that OB stars 
generally span a range in $v$sin$i$ of 100 to 400 \kms.  
Also, de~Jager (1980, p.50) summarizes mean equatorial velocities for 
B1/B2~V stars (the majority of stars in this paper) as $\sim$200~\kms.   
Although all the stars in this paper, and others with good abundance 
analyses, are sharp lined objects, the usual assumption
is that many of these stars are (near) pole-on rotators.

In Fig.~\ref{djltg}, we show the HL00 models for 0 and 200~\kms,
with and without $\mu$-barrier effects in the \logg-log\teff\ diagram.  
We also include our stellar targets, and more B-type stars with 
boron and CNO abundances from the literature, to show their approximate 
mass range and main sequence ages 
(differences between the symbols are discussed below).    
We notice that the model ZAMS is offset from the highest gravity stars. 
This is most likely an artefact of uncertainties in the temperature scale
(discussed below),  
which subsequently affects the determinations of gravity and $\xi$. 

\subsubsection{Boron versus Nitrogen:  Predictions}

Heger \& Langer's (2000) models can reproduce previously unexplained 
observations of massive stars (e.g., various abundance anomalies) and
the models make certain predictions (see their discussion 
of the observational evidence and testable predictions). 
The prediction that concerns us in this paper is the relationship
between boron and nitrogen\footnote{ 
We note that boron versus the ratio N/C is also a good indicator,
but our opinion is that the small depletions in carbon due to mixing 
do not clearly compensate for the added uncertainties in the B-type 
star carbon abundances.}
during main sequence evolution.

Mechanisms that can deplete boron at the surface of stars can 
also enrich the surface in the products of the CN-cycle; material 
that is rich in nitrogen and depleted of carbon, yet where the sum C+N is preserved.  
Extremely deep mixing could also bring ON-cycled material to the surface,
preserving the sum C+N+O.
Thus, in Fig.~\ref{heger}, we show the boron-to-nitrogen relationship 
and the rate of boron depletion predicted by various HL00 models
for main sequence evolution (through core-H burning).    
We show models that include $\mu$-barrier effects for 12~M$_\odot$ 
at five rotational velocities, and for comparison we have included
two 15~M$_\odot$ models at 200~\kms\ and 450~\kms\, as well as the 
10~M$_\odot$ and 15~M$_\odot$ models without the $\mu$-barrier effects.
These figures allow us to examine 
the effects due to rotation rate, mass, age, and mixing efficiencies.

One of the most interesting results is that the boron-nitrogen
relationship is nearly the same for all models.   This occurs because
boron is destroyed throughout most of the star very early in its
main sequence lifetime ($\leq$10$^4$~yrs).  Only the outermost layers
of the stellar interior are cool enough to retain the initial boron
(the outer $\sim$1~M$_\odot$ for 10-20~M$_\odot$ stars).   Meanwhile
during this early ZAMS phase, the CN-cycle begins in the core,
transforms carbon into nitrogen, and mixes this gas into the interior through 
convective overshoot and/or semi-convection.  
As the star ages on the main sequence, rotation will mix the 
CN-cycled gas up through the stellar interior, while also mixing 
the pristine outer layers downwards.  
Fig.~\ref{heger} illustrates that mixing the radiative envelope
of a massive main sequence star will always have the same consequences
for the surface abundances; boron is depleted first, and nitrogen
is enriched only later (also shown in Fliegner \etal 1996). 
Due to the similarity in the structure of the envelopes of
the stars considered, in particular the shallowness of the 
surface layer that contains pristine boron and the depth of the 
layers were nitrogen is enriched, then the same abundance changes 
are predicted independent of all parameters. 

On the other hand, the {\it timescales} for the B-N abundance 
changes do differ significantly between the models.  
Since rotational mixing is less effective for slower rotation 
rates or lower masses, then the abundance changes 
simply take longer to occur.   Furthermore, very few models predict 
high N/C ratios at the end of the main sequence lifetime, 
thus additional processing/mixing after the main sequence phase
(such as the first dredge-up) can be examined through the N/C 
ratios of more evolved stars.
 
We comment once again that mass loss could produce the same B-N abundance
changes as rotation (having the same ZAMS starting conditions as well).
However, the mass loss rates for B-type stars are much smaller than required 
(as discussed earlier), thus favoring rotation for this phenomenon.
Also, boron depletion and nitrogen enrichment are predicted
by mass transfer in a close binary system.   Wellstein (2000) 
and Wellstein \etal (2001) predict changes in surface abundances 
of boron and CNO due to the transfer of nuclear processed matter.   
The current models predict much
larger boron depletions though, and there is always a very strong 
CNO-signature as well.  Therefore, the existence of moderately
boron-depleted stars without nitrogen enrichments is a unique 
signature of rotational mixing effects.

\subsubsection{Boron versus Nitrogen:  Observations}

In the abundance analysis described above, we have adopted
the Gies \& Lambert (1992 = GL92) atmospheric parameters.
Systematic differences between GL92 and other B-type star 
analyses suggest that this temperature scale is too hot
(e.g., Korotin \etal 1999, Cunha \& Lambert 1994 = CL94). 
In fact, GL92 did increase their temperature scale by 3.4\% from 
the photometric determinations.  This increase removed an unexpected
relationship between NLTE CNO abundances and temperature.
However, this relationship appears to have been due to the NLTE
CNO abundances having been determined from Gold (1984) model atmospheres,
instead of the more heavily line-blanketed Kurucz models.   
 
In Table~\ref{bcno}, we have reduced the GL92 temperature scale 
by 3.4\%, and applied their NLTE CNO abundance corrections listed
in their Table~9 (the $\Delta$ values).   
We have also applied a correction to account
for the Gold-Kurucz offsets, as tabulated by CL94 (their Table~10).
(Note that this latter correction is derived from differences in LTE
abundances between Gold and Kurucz models, and we assume the same
will apply to the NLTE abundances, as assumption that should be checked).    
The GL92 temperatures are now in good agreement
with CL94 values for stars in common. 
In the case of \bd, which was not analysed by GL92 or CL94, 
we take the parameters and NLTE CNO results of Vrancken \etal (2000), 
but add 0.4~dex to their published nitrogen value to account for the systematic 
differences they discuss between their analysis and those by GL92 
and Kilian (1992, 1994).  

The GL92 NLTE nitrogen abundances are now in excellent agreement with CL94 
values for stars in common.   Only the hottest star, HD\,36960 shows
a non-negligible nitrogen difference of $\sim$0.1~dex.  However, it is
surprising that the carbon and oxygen abundances often differ between these 
analyses, by up to 0.25~dex.
For carbon, the abundances differ when GL92 include two strong 
\ion{C}{3} lines.    For oxygen, the differences are traceable to 
the adopted $\xi$ values.   CL94 adopt higher $\xi$ values as 
their analysis is more sensitive to $\xi$ due to some stronger lines.  
For consistency, we adopt the corrected GL92 values for CNO 
in Table~\ref{bcno} whenever possible. 

The boron abundances determined from the \ion{B}{3} feature
are not sensitive to these uncertainties in temperature, except
for the hottest and coolest stars that we consider less reliable 
boron indicators (discussed above), and thus that we have excluded 
from Table~\ref{bcno} (including our hot target HD\,34078).
We also note that the uncertainties in \teff\ and/or gravity do
not significantly affect the boron-nitrogen relationship that we 
seek; i.e., the \ion{B}{3} and \ion{N}{2} line abundances react 
similarly to the atmospheric parameters in most cases (only in 
the coolest stars is nitrogen more sensitive to the atmospheric parameters).  
Thus, uncertainties in the parameters cannot induce a B-N anticorrelation.

\subsubsection{Predictions versus Observations}

To compare the boron and nitrogen abundances in B-type stars to the
model predictions, we have gathered abundances from this analysis and 
the literature, corrected them to be on a homogeneous temperature
scale, and applied the known Gold-Kurucz model atmosphere abundance 
corrections (discussed above).  We have also limited the atmospheric 
parameter range of our sample to 18,000~K~$\le$~\teff~$\le$~29,000~K, 
where the \ion{B}{3} line strength is in a plateau, 
and log~g $\ge$3.4 to examine main sequence stars only.  
GL92 showed that their sample of 
B-type stars contained some stars mildly depleted of carbon and 
enriched in nitrogen with no measureable spread in the oxygen abundances.  
The C-N signature is indicative of CN-cycled material in the 
atmospheres of some stars.
There is no evidence for ON-cycled gas in main sequence B-type stars.

In Fig.~\ref{modbn}, we can see that most stars gather 
around the point 12+log(B/H)~=~2.6 and 12+log(N/H)~=~7.7,
consistent with the ISM boron abundance in Orion (discussed above)
and the ISM nitrogen abundance 
(e.g., 12+log(N/H)~=~7.8, Esteban \etal 1998).
Possibly, there is a small intrinsic spread - initial boron may 
be higher (about 2.8) and initial nitrogen lower (about 7.6) -
or perhaps some small depletion by rotation has already occured.  
Boron and nitrogen predictions from HL00 for the 12~M$_\odot$ 
models (with 200 \kms\ and $\mu$-barrier effects) are shown 
in Fig.~\ref{modbn}.    The same model is shown scaled with 
three different initial boron and nitrogen abundances (all initial 
abundances used here are within the observational uncertainties) 
to better compare to the stellar observations. 
Only the predicted $^{11}$B abundances are traced.  The isotope
$^{10}$B is more readily destroyed by protons, such that one 
would assume that the $^{11}$B/$^{10}$B would increase in regions
of partial boron destruction.   However, the HL00 models predict 
a slower depletion rate for $^{10}$B.  Thus, $^{11}$B/$^{10}$B~=~4 
adopted for the ZAMS is predicted to reduce to $\ge$2 in the stellar 
envelope by the end of the main sequence.    
We note that these differences in the boron isotopic 
ratio have a negligible effect on the boron abundance 
determinations in most stars (see Table~\ref{boron-unc}). 
Only two stars have low-boron {\it determinations}, HD\,16582 
(this paper) and HD\,3360 (Proffitt \etal 1999).  
Several others have interesting upper limits that follow 
the boron-nitrogen trend predicted by the rotating models.  
HD\,36591 in Orion is particularly interesting since it shows 
a strong boron depletion with no enrichment in nitrogen;    
the model predictions are most consistent with this
star for the lowest initial nitrogen (=7.6) abundance. 

In Fig.~\ref{modage}, we examine the boron abundances 
with stellar age. 
Three stars in this paper are in young clusters, and 
show boron depletions that are fit well by rotating 
model predictions.   Two of these stars have nitrogen enrichments, 
thus they could also be explained by binary accretion 
(although only one is in a known binary system), 
but the third has unenriched nitrogen.   As previously 
stated, the best interpretation for boron-depletion without 
nitrogen-enrichment is rotational mixing.  This third star 
is HD\,36591, again, and we find that the most extreme 
models, e.g., 20 M$_\odot$, $v_{eq}$=450~\kms, reproduce 
its age, boron, and nitrogen abundances well.   Unfortunately, the
atmospheric parameters for this star make it very unlikely
that it is a 20~M$_\odot$ star (see Fig.~\ref{djltg}).
Perhaps mixing efficiencies above $\mu$-barriers can be 
higher then currently predicted. 

Also seen in Fig.~\ref{modage} is that some B-type stars 
in older clusters have  normal boron abundances.    
There are two possiblities for this.  First, they may 
be true slow rotators (not rapidly rotating stars seen 
pole-on as is usually assumed for sharp-lined objects).   
This is likely true for some stars, e.g., HD\,216916 is 
an eclipsing binary (Pigulski \& Jerzykiewicz 1988), thus 
its low $v$sin$i$ is probably close to its equatorial 
velocity.   Also, some older subgroups of OB associations
have been reported to have an excess of slow rotators
(Guthrie 1984, also see the discussion by Grebel \etal 1992).
Second, some stars may be lower mass objects, 
e.g., the 10 M$_\odot$ models require more time to mix
hotter gas from the interior to the surface, 
thus nitrogen and boron remain near their initial abundances longer. 
Schoenberner and Harmanec (1995) 
determined masses of main sequence B-type stars in binary 
systems, and found that stars in our primary temperature 
range (24000-28000~K) have masses of 10-12~M$_\odot$, 
consistent with Fig.~\ref{djltg}.

In Fig.~\ref{djltg}, we also examine the boron (from \ion{B}{3} only) 
and nitrogen abundances on a \teff-gravity diagram to examine the abundance
distributions across the main sequence.    The stellar data is 
divided into three groups: normal stars (B$>$2.2 and N$\le$7.8), 
boron depleted stars (B$\le$2.2 and N$\le$7.8), 
and nitrogen rich stars (B$\le$2.2 and N$>$7.8).    
As expected, there are no N-rich and B-normal stars.
Evolution tracks from HL00 are shown for 8-20~M$_\odot$
through the core-H burning phase.    
In general, the three groups of stars have the same range 
in \teff\ and gravity, and thus in age and mass, suggesting 
that their differences are due to another parameter, 
like rotation rate.  
It is interesting that most of the B-depleted/N-normal stars
are near the ZAMS, as is predicted for rapid rotators.
Also, one might argue that among the most massive stars, more
of them show B-depletions; this would be consistent 
with the models if the rotation rates are similar throughout. 
 
Finally, it is pleasing that \bd, the star with the 
lowest surface gravity of our sample, has apparently 
retained its initial complement of boron and has unenriched nitrogen.    
This star is in an eclipsing binary system, 
which implies its $v$sin$i$ value is very close to its current 
rotational velocity.   Thus \bd\ is probably a slow rotator today, 
and was probably a slow rotator on the ZAMS (although it may have 
spun down somewhat from its ZAMS rotation rate).
Only one other star shows unenriched nitrogen and a lower 
surface gravity, HD\,30836, but PQ01 show that this
star has depleted boron, suggestive of rotational mixing.
(This star is in Ori OB1, but it is not in Fig.~\ref{modage} 
since its subassociation, thus age, is not clear).

\subsection{PG\,0832+676}

Our determination of the metallicity of PG\,0832+676 is 
[Fe/H]~=~$-$0.88 $\pm$0.10, which compares with 
[Fe/H]~=~$-$0.51 given by Hambly \etal (1996) from 
a single optical Fe\,{\sc iii} line in a differential
analysis with respect to the Galactic B-type star HR~1886.
Combining our [Fe/H] with the mean $\alpha$-element 
(Mg, Si, and S) abundance from Hambly \etal, we obtain
[$\alpha$/Fe]~=~0.5, which is the expected value for a star
of this metallicity to within the errors of measurement
(e.g., [O/Fe]~=~0.3 for [Fe/H]~=~$-$0.9, as for the mildly
metal-poor clusters M4 and M71, Ivans \etal 1999, 
Sneden \etal 1994, although the O/Fe ratio in metal-poor
stars is currently controversial, c.f., Lambert 2000).

Relative to [Fe/H]~=~$-$0.88, PG\,0832+676 is enriched in 
carbon, nitrogen, and oxygen, according to Hambly \etals results: 
[C/Fe] $\simeq$ [N/Fe] $\simeq$ [O/Fe] $\simeq$ 0.7 to 0.8.
This O/Fe ratio is somewhat larger than expected for a normal 
metal-poor star in the halo (see $\alpha$/Fe ratio comments above),
and carbon and nitrogen appear enriched relative to normal stars.
These results are consistent with the identification of PG\,0832+676
as a post-AGB star with carbon and oxygen added in the AGB phase, and nitrogen
enhanced by the earlier first dredge-up.   Certainly this
scenario is consistent with our non-detection of boron.
Our limit of 12+log(B/H)~$\leq$~0.60 is much less than
the expected initial abundance of 12+log(B/H)=1.9 at 
[Fe/H]~=~$-$0.9 (based on the analysis of 14 cool 
dwarfs by Cunha \etal 2000a). 

Thus, in summary, the chemical pattern in PG\,0832+676, 
low [Fe/H], high [$\alpha$/Fe] and [CNO/Fe], very low [B/H], 
suggests that this star is a post-AGB star,
in agreement with Hambly \etals (1996) conclusion that PG\,0832+676
is a highly evolved star.

\section{Conclusions}

Boron in hot stars, like lithium in cool stars, is shown to be 
a tracer of some of the various processes affecting a star's 
surface composition that are not included in the standard 
models of stellar evolution.    If the initial boron abundances 
of local hot stars are similar from star-to-star and 
association-to-association, then the large spread in boron 
abundances, a factor of at least 30 across our sample,  
shows that boron abundances are a clue to unravelling the 
non-standard processes that affect young hot stars.   In this
paper, we have focussed on the role of rotationally-induced mixing.

Models of stars with masses near 10 $M_\odot$ (Heger \& Langer 2000)
show that rotationally-induced mixing during main
sequence evolution can reduce the surface boron 
abundance and increase surface nitrogen. 
A signature of rotationally-induced mixing is that the 
initial decline of boron precedes an observable change in nitrogen. 
The correlation between the decline of boron and the rise in nitrogen 
is almost independent of rotational velocity and mass of the 
star.  On the other hand, the extent and rate of change 
of the abundances are dependent on both velocity and mass 
(Fig.~\ref{heger}). 

Our results confirm Proffitt \etals (1999) discovery that 
boron may be quite severely depleted in otherwise normal B-type 
stars; 6 of the 8 stars in our program show a reduced boron abundance. 
That boron depletions are not a rare occurrence is consistent 
with Proffitt \& Quigley's (2001) survey of {\it IUE} spectra.   
Boron depletions are attributed to rotationally-induced mixing 
(although N-rich stars could also be explained by binary mass 
transfer).  The correlation between the boron and nitrogen abundances follows 
the predicted trend quite well (Fig.~\ref{modbn}).

Our use of stars from various OB associations allows us to investigate 
the rate of change of the surface abundances.  Stars showing a normal 
boron abundance are found at all the investigated ages; this is 
consistent with predictions for rotationally-induced mixing 
for low equatorial velocities (100\,$-$\,200 \kms).
The observed $v$sin$i$ velocities are considerably smaller than 
these limits, but we assume most stars are rapidly rotating yet 
seen pole-on.  Two stars appear to have depleted boron at a far
faster rate than expected; HD\,36591 and HD\,205021 are very young 
stars in our program with no detectable boron.   In Fig.~\ref{modage}, 
they appear close to the locus for stars of 20 $M_\odot$ rotating at 
450 \kms, but their locations in Fig.~\ref{djltg} imply that the stars 
are very early main sequence stars with masses of about 
12\,$-$\,13 $M_\odot$.  This discrepancy may indicate that
rotationally-induced mixing has been underestimated at high rotational
velocities or that it is more efficient then expected at lower masses.

The spectral window around the \ion{B}{3} resonance lines proves
well suited for a determination of the abundance of the iron-group
elements. For the local B-type stars in our sample, we show that they 
have an approximately solar metallicity. For the halo B-type star
PG\,0832+676, we find a low metallicity and an absence of boron which
suggest the star is probably a post-AGB star.

\begin{figure}
\plotone{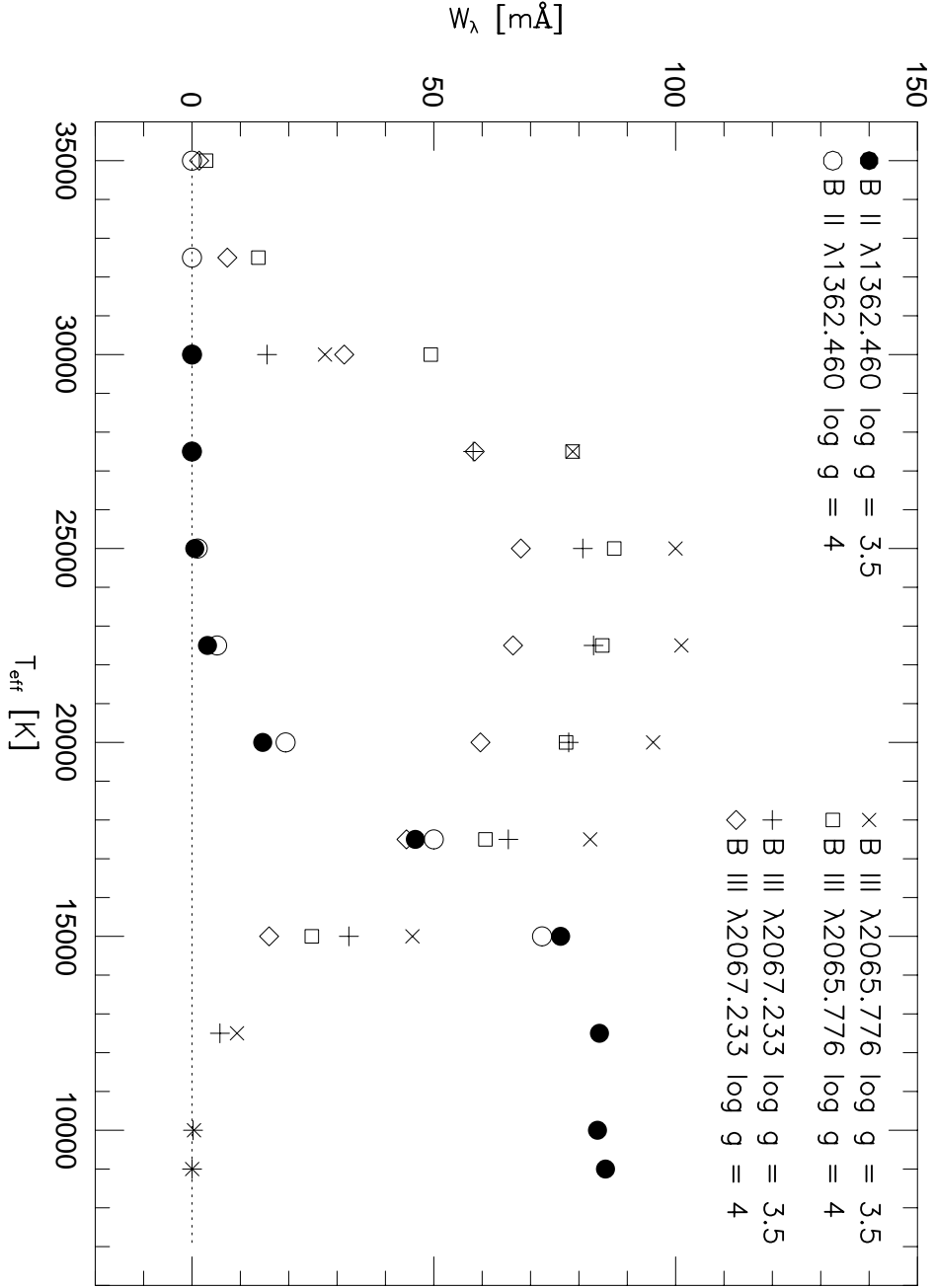}
\caption{Predicted equivalent widths for boron lines in B-type stars,
from NLTE calculations.   The \ion{B}{3}~2065.8 line strength plateaus
between 18,000~K and 29,000~K.   At hotter temperatures,
an increase in \teff\ of only 5~\% reduces the predicted 
equivalent width by 50~\%, or 0.4~dex in the boron abundance.
\label{beqw}}
\end{figure}

\begin{figure}
\plotone{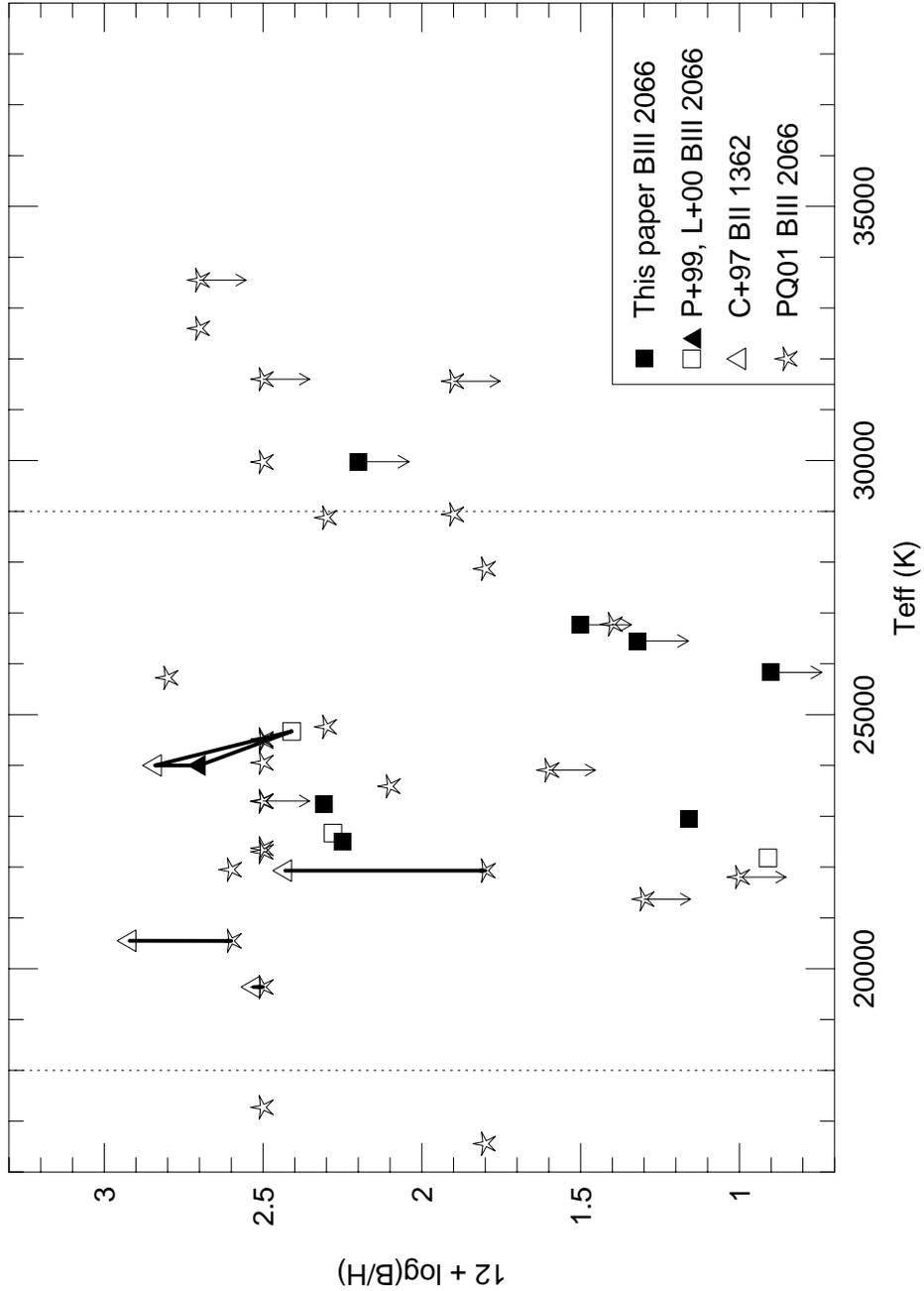}
\caption{Boron abundances versus temperature.   
Temperatures from Gies \& Lambert (1992) have 
been reduced by 3.4\% (see text).   This plot 
shows that there is no significant trend in 
the \ion{B}{3}~2065.8\,\AA\ abundances, although we 
note the upper and lower temperature limits 
where the \ion{B}{3}~2065.8\,\AA\ line yields the most 
reliable abundances.   {\it Thick lines} connect 
boron abundances found for the same stars from 
different analyses.
\label{bt}}
\end{figure}

\begin{figure}
\plotone{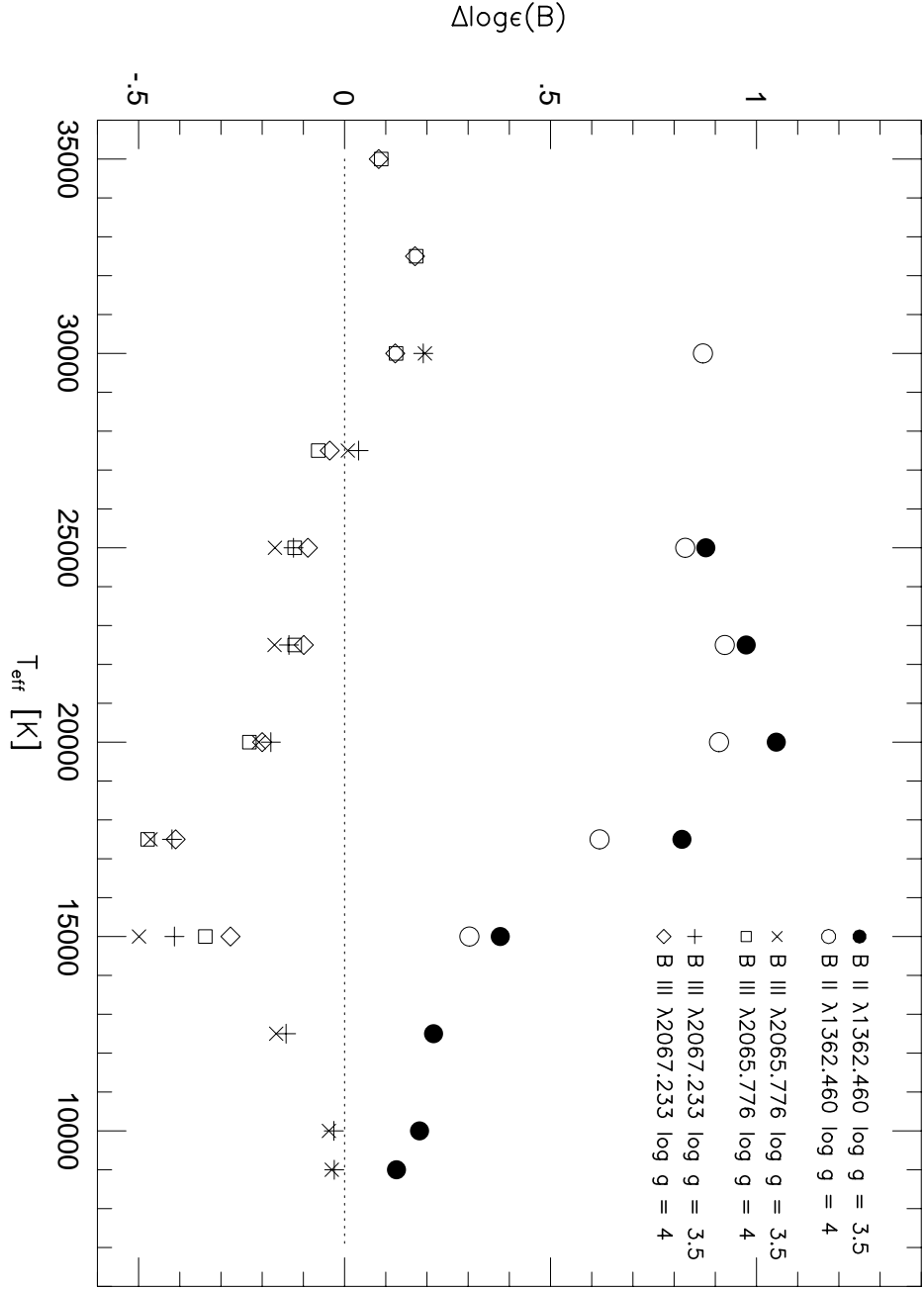}
\caption{NLTE corrections to UV boron lines.   These calculations 
are identical to those discussed in Cunha \etal (1997); here, 
we extended the grid to hotter temperatures (35,000~K) 
for the \ion{B}{3} features.  The change in orientation 
of the \ion{B}{3} NLTE correction in the hotter models appears to 
be due to the much weakened boron lines being dominated by 
overionization effects (as opposed to line transition rates, which 
dominate at lower temperatures). 
\label{nlte}}
\end{figure}

\begin{figure}
\epsscale{.9}
\plotone{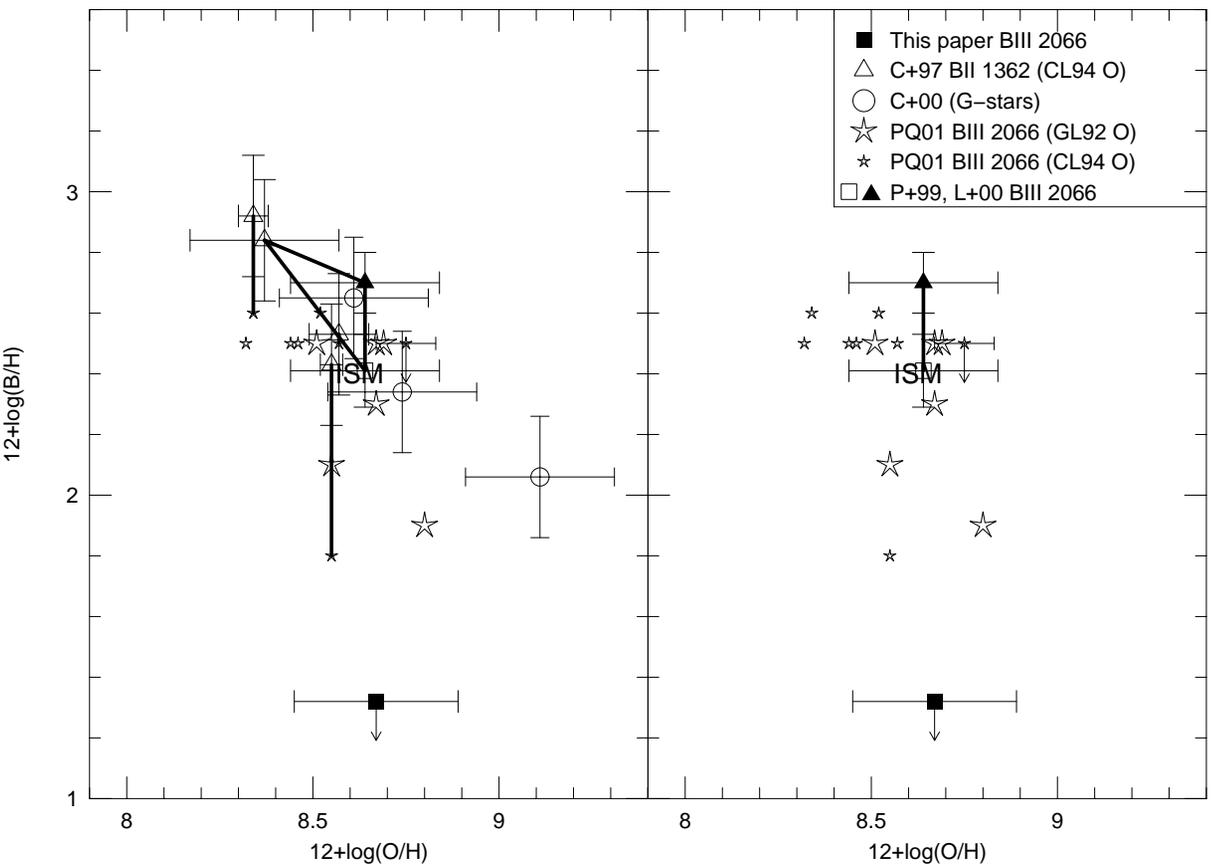}
\caption{Boron versus oxygen abundances in Orion stars.   
Left panel shows all stars with both boron and oxygen
abundances, including B-type and G-type stars, and the
interstellar abundances.   Left panel also shows the effect 
of replacing \ion{B}{2} abundances with those from the 
\ion{B}{3} feature ({\it thick lines}).
Right panel shows only the \ion{B}{3} abundances in B-type 
stars.  The \ion{B}{3} abundances do {\it not} confirm the 
B-O anti-correlation (see text for further discussion), 
although the PQ01 {\it IUE} data points have large 
uncertainties that are not shown for clarity. 
\label{boxy}}
\end{figure}

\begin{figure}
\epsscale{.9}
\plotone{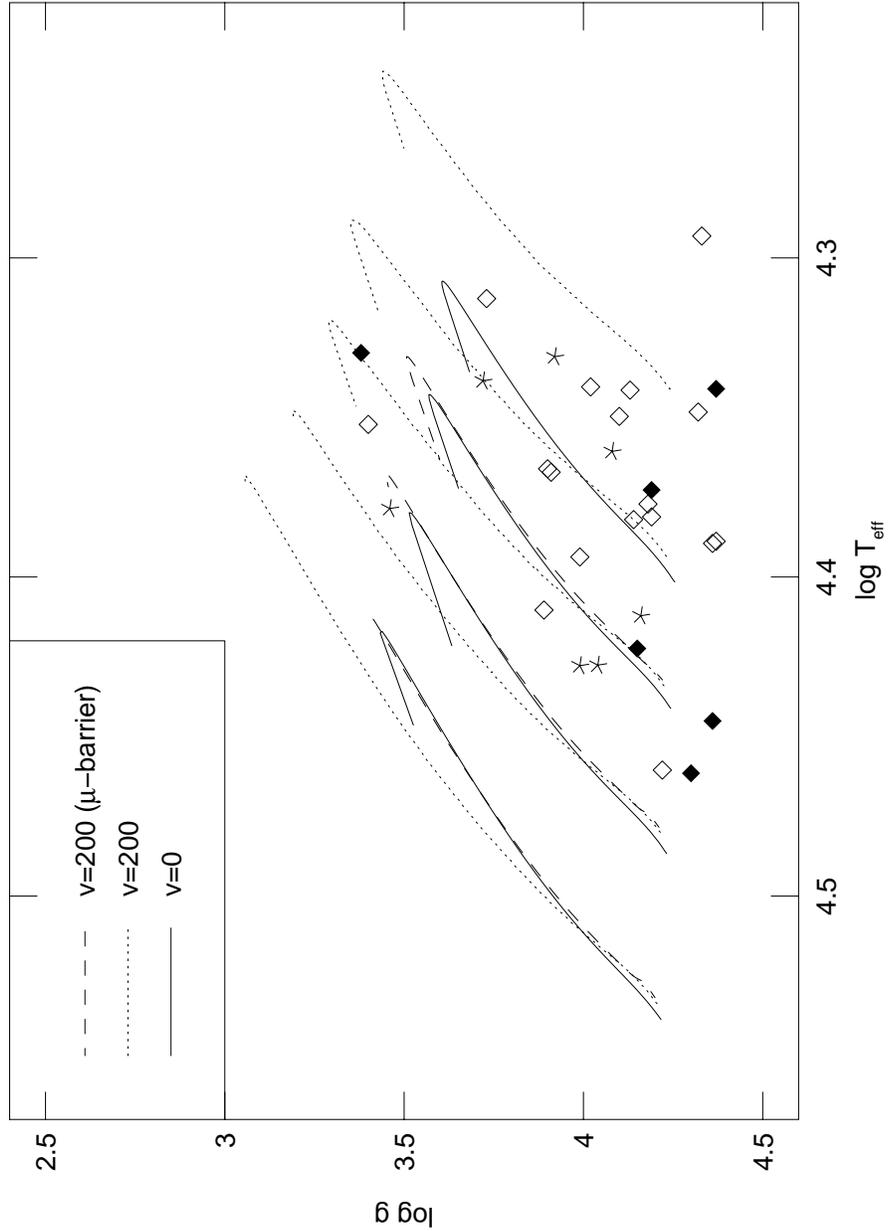}
\caption{Gravity versus \teff\ for B-type main sequence stars 
in Table~\ref{bcno}.  {\it Open diamonds} mark normal stars 
(B$>$2.2 and N$\le$7.8), {\it filled diamonds} show boron depleted 
stars (B$\le$2.2 and N$\le$7.8), and {\it asterisks} denote the 
N-rich stars (B$\le$2.2 and N$>$7.8).   Evolution tracks are
from Heger \& Langer (2000) for stars of 8, 10, 12, 15,
and 20 M$_\odot$, two rotation rates, and considering 
$\mu$-barrier effects, through the H-core burning phase.
This plot shows the mass and age range of the B-type stars
examined in this paper, as well as the distribution of boron
depletions and nitrogen enrichments on the main sequence.
\label{djltg}}
\end{figure}

\begin{figure}
\epsscale{.83}
\plotone{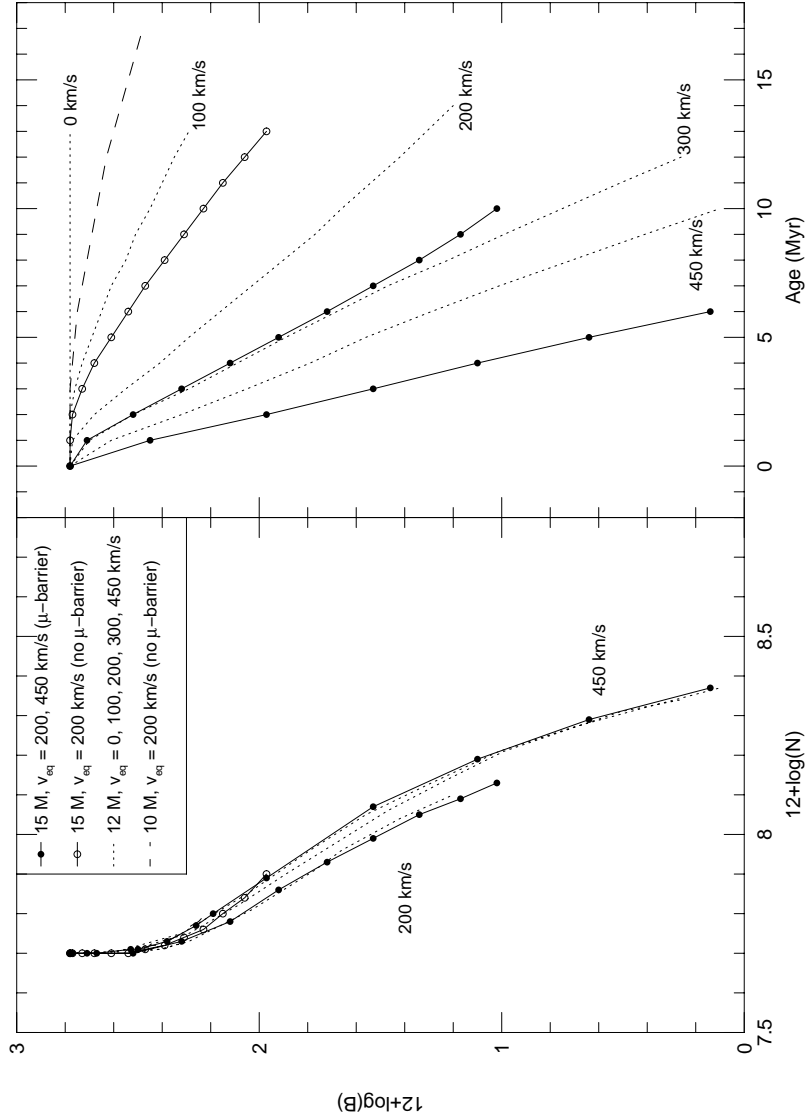}
\caption{Predictions of boron versus nitrogen (left panel) and 
boron versus main sequence age (right panel) from the rotating 
stellar evolution models from Heger \& Langer (2000). 
Predictions for a 12~M$_\odot$ model with five rotational
velocities are shown ({\it dotted lines}, velocities are 
marked in the right panel).
For comparison, predictions for 
a 15~M$_\odot$ model with rotational velocities 
of 200 and 450~\kms\ are shown ({\it solid lines}).
Additionally, 10 and 15~M$_\odot$ models, rotating at 200~\kms, 
and that ignore the effects of $\mu$-barriers are shown. 
In the left panel, it is striking that the boron versus nitrogen 
relationship is nearly the same for various masses, 
rotation rates, and efficiency of mixing with/without 
$\mu$-barriers, although some tracks end before significant
abundance changes.  However, in the right panel, it can be 
seen that the {\it rate} of boron depletion is sensitive 
to these parameters.
\label{heger}}
\end{figure}

\begin{figure}
\plotone{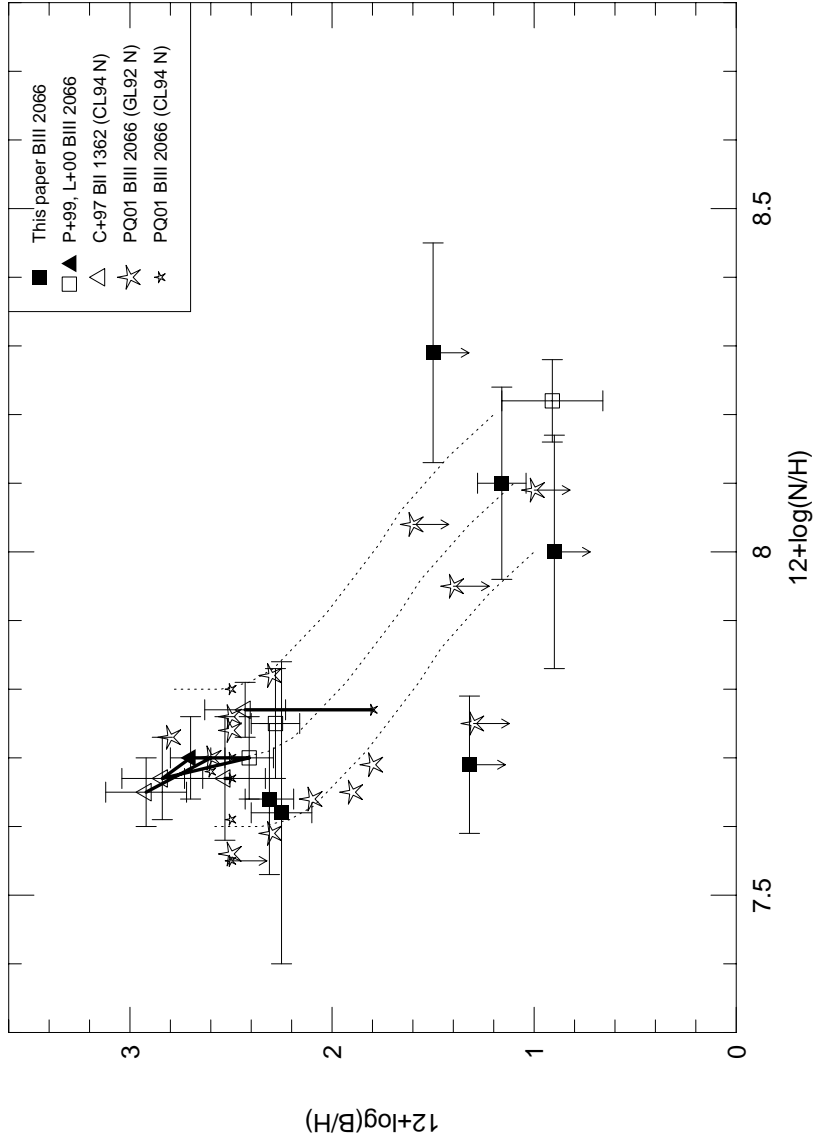}
\caption{Boron versus nitrogen in main sequence B-type stars, 
with predictions for the 12~M$_\odot$ models at 200~\kms\ 
through the H-core burning phase, and three sets of initial 
abundances [(B$_i$, N$_i$) = (2.6, 7.6), (2.8, 7.7), and 
(2.8, 7.8), all within range of the interstellar and ZAMS 
stellar abundances]. 
The stellar abundances plotted are those listed in 
Table~\ref{bcno}.  {\it Thick lines} connect boron 
abundances for the same stars from different analyses.
The observations are in good agreement with the predictions,
suggesting that rotational mixing is significant on
the main sequence in some B-type stars. 
\label{modbn}}
\end{figure}

\begin{figure}
\epsscale{.83}
\plotone{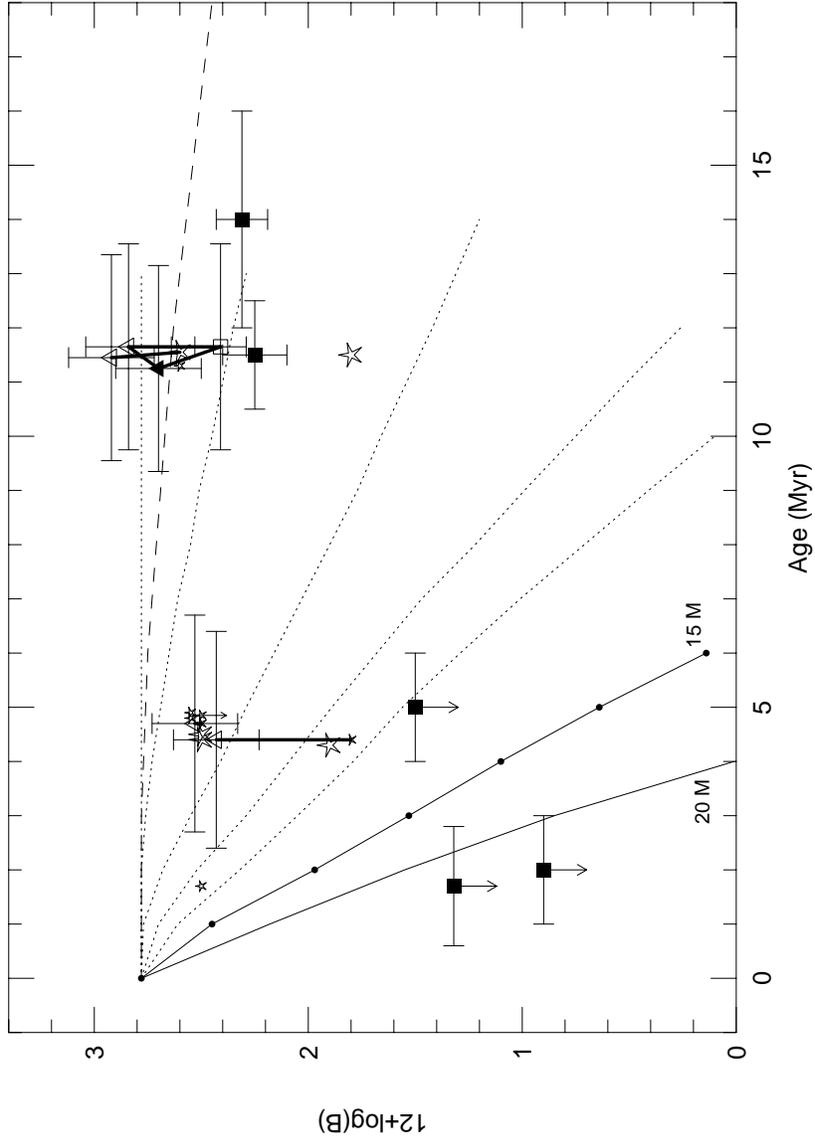}
\caption{Boron versus cluster ages (see Table~\ref{varbin}
for ages and references) for B-type stars in associations.  
Stellar data symbols are the same as in Fig.~\ref{modbn}.
Tracks are the same as in Fig.~\ref{heger} (some tracks
are not shown here for clarity).  One additional track
is included to show the predictions for a 20~M$_\odot$ star 
with $\mu$-barrier effects, rotating at 450~\kms
({\it solid line}).
Three boron depleted young stars (HD\,36591, HD\,50707, HD\,205021) 
are fit well by the rotating model predictions, although the
most boron depleted stars are best fit by the rapidly-rotating 
20~M$_\odot$ model, which is likely too massive for these
stars (see Fig.~\ref{djltg}).   
Many of the older stars in this sample that do not show
boron depletions are less massive (e.g., the 10~M$_\odot$ track,
{\it dashed line}, fits them well), though some are likely 
to be true slow rotators as well.  
\label{modage}}
\end{figure}

Further insights into rotationally-induced mixing will require
STIS spectra of additional rapidly rotating stars.   A few
stars examined by Proffitt \& Quigley (2001) have low boron upper-limits
and unenriched nitrogen, and are ideal targets for follow-up studies. 
Additional stars that are sharp-lined (low $v$sin$i$), but 
rapidly rotating, are also suitable; however, in the absence 
of a way to extract the angle of inclination, information 
about abundances and rotation will require a statistical treatment.   
We also note that stars hotter than the present sample are not 
appropriate; the \ion{B}{3} lines become too weak (see Fig.~\ref{beqw}) 
and boron, present as the He-like ion, is lost to spectroscopic scrutiny.

\acknowledgments
Support for proposal GO\#07400 was provided by NASA through a 
grant from the Space Telescope Science Institute, which is
operated by the Association of Universities for Research in Astronomy, Inc., 
under NASA contract NAS 5-26555.  KAV, ML, and AB would also like to
acknowledge research support from Macalester College and the Luce Foundation 
through a Clare Boothe Luce Professorship award.   
KAV thanks Grace Mitchell and Claus Leitherer for help with the STIS data 
reductions and for STScI visitor funds.  Many thanks to Charles Proffitt 
for helpful discussions, comments on the manuscript, and making his {\it IUE} 
data available for inspection.   We also thank Katia Cunha for 
helpful discussions.

%\appendix
%\section{Appendix}

%% Input the Bibliography, bibliography 

%% Input the Tables, TABLES, tables
% "deluxetable" .
\clearpage
\begin{deluxetable}{lccrrr}
%\footnotesize
\tablecaption{HST STIS Observing Information for Galactic B-stars \label{obs}}
\tablewidth{0pt}
\tablehead{
\colhead{Star} & \colhead{V} & \colhead{Grat/Slit} & 
\colhead{Exposure(s)} & \colhead{Date} & \colhead{S/N}
} 
\startdata
BD+56\,576   &  9.38 & E230M     &   1680s at $\lambda_c$2124 & 11 FEB 99 & 100 \nl
------       &       & 0.2x0.2   &  +1280s at $\lambda_c$1978 & \nl 
------       &       &           &  +1509s at $\lambda_c$2269 & \nl 
HD\,16582    &  4.07 & E230H     &   1634s at $\lambda_c$2113 & 23 JAN 99 & 90 \nl
------       &       & 0.2x0.05ND & +1140s at $\lambda_c$2063 & \nl
------       &       &           &  +1140s at $\lambda_c$2063 & \nl
------       &       &           &  +1368s at $\lambda_c$2013 & \nl
------       &       &           &  +1328s at $\lambda_c$2013 & \nl
HD\,34078    &  6.00 & E230H     &    432s at $\lambda_c$2063 & 15 MAR 00 & 50 \nl
------       &       & 0.1x0.03  &   +432s at $\lambda_c$2013 & \nl
HD\,36591    &  5.34 & E230M     &    432s at $\lambda_c$2124 & 09 FEB 99 & 100 \nl
------       &       & 0.2x0.05ND &  +432s at $\lambda_c$2124 & \nl 
------       &       &           &   +432s at $\lambda_c$2124 & \nl 
------       &       &           &   +432s at $\lambda_c$1978 & \nl 
------       &       &           &   +432s at $\lambda_c$1978 & \nl 
------       &       &           &   +432s at $\lambda_c$2269 & \nl 
------       &       &           &   +432s at $\lambda_c$2269 & \nl 
HD\,50707    &  4.83 & E230M     &    216s at $\lambda_c$2124 & 06 NOV 98 & 75 \nl
------       &       & 0.2x0.05ND &  +216s at $\lambda_c$1978 & \nl 
------       &       &           &   +216s at $\lambda_c$2269 & \nl 
HD\,205021   &  3.23 & E230H     &    900s at $\lambda_c$2063 & 19 FEB 99 & 100 \nl
------       &       & 0.2x0.05ND &  +900s at $\lambda_c$2063 & \nl
------       &       &           &   +684s at $\lambda_c$2013 & \nl
------       &       &           &   +684s at $\lambda_c$2113 & \nl
------       &       &           &   +684s at $\lambda_c$1963 & \nl
HD\,216916   &  5.59 & E230M     &   1260s at $\lambda_c$2124 & 16 DEC 98 & 130 \nl
------       &       & 0.2x0.05ND &  +262s at $\lambda_c$2124 & \nl
------       &       &           &   +338s at $\lambda_c$2124 & \nl
------       &       &           &   +888s at $\lambda_c$1978 & \nl
------       &       &           &   +972s at $\lambda_c$2269 & \nl
PG\,0832+676 & 14.15 & E230M     &   3360s at $\lambda_c$2124 & 08 DEC 98 & 100 \nl 
------       &       & 0.2x0.2   &  +3360s at $\lambda_c$2124 & CVZ \nl 
------       &       &           &  +3360s at $\lambda_c$2124 &  \nl 
------       &       &           &  +3300s at $\lambda_c$2124 &  \nl 
\enddata
\end{deluxetable}
\clearpage

% "deluxetable" .
\clearpage
\begin{deluxetable}{llllrrr}
%\footnotesize
\tablecaption{Variability and Binary Information from the Literature \label{varbin}}
\tablewidth{0pt}
\tablehead{
\colhead{Star} & \colhead{Name} & \colhead{Variable} &  \colhead{Binary} &  
\colhead{Cluster}  & \colhead{Cl. Age} & \colhead{REFS} \\[.2ex]
\colhead{} & \colhead{} & \colhead{} & \colhead{} &
\colhead{} & \colhead{(Myr)} & \colhead{} 
}
\startdata
BD+56\,576 & \nodata & \nodata     & eclipse & $\chi$ Per  & 11.5 & KP97, O37, T+84    \nl
HD\,16582  & $\delta$ Cet & $\beta$ Cep & \nodata & Cas-Tau     & $\sim$50 & H+94, C63, deZ+99 \nl 
HD\,34078  & AE Aur & var & visual? & Ori OBI & \nodata & G87 (run-away)  \nl
HD\,36591  & \nodata & var & visual  & Ori OBIb & 1.7 $\pm$1.1 & O77, BV97, WH78, B+94 \nl
HD\,50707  & 15 CMa & $\beta$ Cep & \nodata & Coll 121    & $\sim$5 & H+94, deZ+99   \nl
HD\,205021 & $\beta$ Cep & $\beta$ Cep & spect   & Cep OB1?    & 2 $\pm$1 & T+97, F69, M+95  \nl
HD\,216916 & 16 Lac & $\beta$ Cep & eclipse & Lac OB1 & 12-16 & PJ88, JP99, F69, B91, deZ+99 \nl
PG\,0832+676 & \nodata &  \nodata   & \nodata & Halo        & \nodata          & \nodata \nl
\enddata
\tablecomments{
References are
B91 = Blaauw 1991,
BV97 = Brown \& Verschueren 1997,
B94 = Brown \etal 1994,
C63 = Crawford 1963,
deZ+99 = de~Zeeuw \etal 1999,
F69 = Fitch 1969,
G87 = Gies 1987,
H+94 = Heynderickx \etal 1994,
JP99 = Jerzykiewicz \& Pigulski 1999,
KP97 = Krzesi\'nski \& Pigulski 1997,
M+95 = Massey \etal 1995,
O77 = Olsen 1977,
O37 = Oosterhoff 1937,
PJ88 = Pigulski \& Jerzykiewicz 1988,
T+84 = Tapia \etal 1984,
T+97 = Telting \etal 1997,
WH78= Warren \& Hesser 1978.}
\end{deluxetable}
\clearpage

% "deluxetable" .
\clearpage
\begin{deluxetable}{lcrrcll}
%\footnotesize
\tablecaption{Atmospheric Parameters from the Literature \label{atms}}
\tablewidth{0pt}
\tablehead{
\colhead{Star} & \colhead{SpType} & \colhead{\teff} & \colhead{\logg}  & 
\colhead{ $v$\,sin\,$i$}  & \colhead{ $\xi$} &
\colhead{REF}   \\[.2ex]
\colhead{} & \colhead{} & \colhead{(K)} & \colhead{} &
\colhead{ (\kms)} & \colhead{(\kms)} & \colhead{}   \\[.2ex]
\colhead{} & \colhead{} & \colhead{} & \colhead{} & 
\colhead{} & \colhead{(NLTE)} & \colhead{}  } 
\startdata
BD+56\,576 & B2III & 22500 & 3.40 & $<$50\tablenotemark{a} & 9 & V+00  \nl
------     & B2III & 21500 & 3.6  & \nodata & 12 &  L88  \nl
HD\,16582  & B2IV & 23750 & 4.08 & 15 & 3.8 $\pm$5.1 & GL92  \nl
HD\,34078  & O9.5V & 31420 & 4.07 & 27 & 8.0 $\pm$0.6 & GL92 \nl  
HD\,36591  & B1IV & 27380 & 4.15 & 11 & 4.4 $\pm$3.7 &  GL92  \nl
------     & B1V & 26330 & 4.21 & \nodata   & 9     &  CL94  \nl
HD\,50707  & B1IV & 27710 & 4.04 & 39 & 4.2 $\pm$4.6 &  GL92  \nl
HD\,205021 & B1IV & 26740 & 4.16 & 28 & 3.3 $\pm$3.7 &  GL92  \nl
HD\,216916 & B2IV & 24050 & 3.90 & 13 & 3.6 $\pm$4.6 &  GL92  \nl
PG\,0832+676 & B1V & 23000 & 3.7 &  1 & {\it 7\,\tablenotemark{b}} & H96 \nl 
\enddata
\tablecomments{References are 
GL92 = Gies \& Lambert (1992), 
V+00 = Vrancken \etal (2000),
H96 = Hambly \etal (1996), 
CL94 = Cunha \& Lambert (1994),
L88 = Lennon \etal (1988).
For BD+56\,576 and HD\,36591, the parameters from V+00 and GL92 were adopted,
respectively. }
\tablenotetext{a}{For BD+56\,576, $v$sin$i$~=~17~\kms\ from our spectrum 
syntheses.}
\tablenotetext{b}{$\xi$ in PG\,0832+676 from a LTE analysis.} 
\end{deluxetable}
\clearpage

% "deluxetable" .
\clearpage
\begin{deluxetable}{lrrrrrrc}
%\footnotesize
\tablecaption{Boron and Iron-Group Abundances from STIS Spectroscopy 
              \label{abus}}
\tablewidth{0pt}
\tablehead{
\colhead{Star} & \colhead{RV} & 
\colhead{$\xi_{Ma}$} & \colhead{$\xi$} &
\colhead{ [$M$/H]} & \colhead{ log($B$/H)}  & 
\colhead{log($B$/H)} & \colhead{ log(Mn~III) }   \\[.2ex] 
\colhead{} & \colhead{ } & 
\colhead{} &  \colhead{ } &  
\colhead{} & \colhead{LTE} & \colhead{NLTE} &
\colhead{ $\lambda$2065.9} 
} 
\startdata
BD+56\,576 &     1 & 20  & 4 & $-$0.16 $\pm$0.17 &      2.43  & 2.25      & 5.24 \nl 
HD\,16582  &    12 & 12  & 2 & $-$0.15 $\pm$0.15 &      1.27  & 1.16      & 5.31 \nl 
HD\,34078  &    56 & 7   & 3 & $-$0.2 $\pm$0.2   & $\le$2.0   & $\le$2.2  & 5.1  \nl
HD\,36591  &    31 & 18  & 2 &   +0.02 $\pm$0.12 & $\le$1.45  & $\le$1.38 & 5.43 \nl
HD\,50707  &    28 & 20  & 6 &   +0.1 $\pm$0.1   & $\le$1.5   & $\le$1.5  & 5.4  \nl 
HD\,205021 & $-$29 & 12  & 2 & $-$0.16 $\pm$0.23 & $\le$1.00  & $\le$0.90 & 5.00 \nl 
HD\,216916 & $-$18 & 19  & 1 &   +0.16 $\pm$0.18 &      2.44  & 2.31      & 5.33 \nl
PG\,0832+676 & $-$68 & 20 & 2 & $-$0.88 $\pm$0.10 & $\le$0.75 & $\le$0.60 & 4.65 \nl 
\enddata
\tablecomments{Abundances have been determined from spectrum syntheses
using the model atmosphere parameters listed here and in Table~\ref{atms}. 
Radial velocities, $\xi$, and $\xi_{Ma}$ (macroturbulence)
values are determined from the iron-group features.
We estimate $\Delta$RV and $\Delta\xi_{Ma}\,\sim$2~\kms\
based on line profile shapes, and $\Delta\xi\le$1~\kms.
The \ion{B}{3} $\lambda$2065.8 and \ion{Mn}{3} $\lambda$2065.9 abundances
were allowed to vary independently for the best-fit syntheses, 
thus we report the \ion{Mn}{3} 2065.9 line abundance here.}
\end{deluxetable}
\clearpage

% TABLE3.TEX -- Sample table 3.

% In this example we could not use the \phs command in place of the 
% \phantom{-} command since \phs is already in math mode.  We could of
% course have reformatted the table, but chose not to do so.

\begin{deluxetable}{lll}
\tablewidth{20pc}
\tablecaption{Iron-Group Wavelength Offsets \label{offset-lines}}
\tablehead{
\colhead{Elem} & \colhead{$\lambda$(KUR)} & \colhead{ $\lambda$(NEW) } }
\startdata
\ion{Mn}{3} & 2048.949 &   2048.909  \\
\ion{Mn}{3} & 2049.357 &   2049.314  \\ 
\ion{Mn}{3} & 2049.682 &   2049.663  \\
\ion{Mn}{3} & 2052.739 &   2052.673  \\
\ion{Mn}{3} & 2063.337 &   2063.397  \\
\ion{Mn}{3} & 2065.886 &   2065.892\tablenotemark{^*} \\
\\
\ion{Fe}{3} & 2050.743 &  2050.738  \\
\ion{Fe}{3} & 2052.271 &  2052.250  \\
\ion{Fe}{3} & 2053.524 &  2053.508  \\
\ion{Fe}{3} & 2054.492 &  2054.475  \\
\ion{Fe}{3} & 2055.863 &  2055.848  \\
\ion{Fe}{3} & 2056.152 &  2056.147  \\
\ion{Fe}{3} & 2057.059 &  2057.052  \\
\ion{Fe}{3} & 2058.209 &  2058.199  \\
\ion{Fe}{3} & 2058.566 &  2058.558  \\
\ion{Fe}{3} & 2059.692 &  2059.678  \\
\ion{Fe}{3} & 2064.980 &  2065.030  \\
\ion{Fe}{3} & 2068.896 &  2068.983  \\
\ion{Fe}{3} & 2070.976 &  2070.976  \\
\ion{Fe}{3} & 2076.322 &  2076.309  \\
\\
\ion{Ni}{3} & 2045.412 &  2045.387  \\ 
\enddata
\tablecomments{Our line list originates from 
Kurucz (1998, CD18), with wavelengths for eight 
\ion{Fe}{3} lines updated from Proffitt \etal (1999) 
and 64 others updated from Ekberg (1993; wavelengths 
and $gf$-values).  In this table, we note additional 
adjustments to some line wavelengths based on our 
preliminary syntheses of sharp-lined stars in our sample.
One line from Ekberg, 2070.976~\AA, was moved back to the 
original Kurucz wavelength, but the new $gf$-value retained.} 
\tablenotetext{^*}{New wavelength from Proffitt \etal (1999), 
who also suggest log$gf$=$-$0.241.   This feature
is blended with the \ion{B}{3} 2065.8 line. }
\end{deluxetable}
\clearpage

\begin{deluxetable}{lrrrrrrrrr}
\tablewidth{0pc}
\tablecaption{Iron-group Abundance Results \label{metals}}
\tablehead{
\colhead{$\lambda$} & \colhead{Element(s)} &
\colhead{ [$M$/H] } & \colhead{ [$M$/H] } & 
\colhead{ [$M$/H] } & 
\colhead{ [$M$/H] } & \colhead{ [$M$/H] } & \colhead{ [$M$/H] } &  
\colhead{ [$M$/H] } & \colhead{ [$M$/H] } \\[.2ex]
\colhead{(\AA)} & \colhead {} &
\colhead{BD56} & \colhead{16582} & \colhead{34078} & \colhead{36591} & 
\colhead{50707} & \colhead{205021} & \colhead{216916} & \colhead{PG0832}}
\startdata
2045.39 & \ion{Ni}{3} & \nodata & $-$0.07 & \nodata  & +0.09   & 
	\nodata & \nodata & +0.16 & \nodata \\

2048.91 & \ion{Mn}{3} & 0.00   & +0.03  & {\it +0.5}  &  0.00 & 
	+0.2\tablenotemark{d} & $-$0.05 & +0.08 & $-$0.96 \\

2049.37 & \ion{Fe}{3}+\ion{Mn}{3} & 0.00   & $-$0.24 & $-$0.3\tablenotemark{a}  & +0.05 & 
	+0.2\tablenotemark{d} & +0.25\tablenotemark{g} & +0.20 & $-$0.90\tablenotemark{i} \\

2049.66 & \ion{Mn}{3} & \nodata & {\it +0.34} & $-$0.3\tablenotemark{a} & +0.05 & 
	+0.2\tablenotemark{d} & +0.25\tablenotemark{g} & {\it +0.77} & {\it +0.09} \\  

2051.85 & \ion{Fe}{3}+\ion{Fe}{4} & $-$0.20 & $-$0.30 & \nodata  & +0.25   & 
	\nodata & \nodata & \nodata & \nodata \\

2052.25 & \ion{Fe}{3} & $-$0.20 & $-$0.25 & \nodata  & +0.13   & 
	\nodata & \nodata & +0.35 & $-$0.93 \\

2053.51 & \ion{Fe}{3} & {\it $-$0.58} & $-$0.35 & $-$0.4 & {\it $-$0.48} & 
	\nodata & 0.00 &  +0.10 & $-$0.98 \\

2054.56 & \ion{Fe}{3}x3    & $-$0.24 & $-$0.06 & $-$0.4  & 0.00   & 
	0.0 & +0.25 & +0.19 & $-$0.84 \\

2055.85 & \ion{Fe}{3} & \nodata & {\it +0.25} & $-$0.2\tablenotemark{b} & +0.08  & 
	+0.2\tablenotemark{e} & \nodata & \nodata & {\it $-$0.24} \\

2056.15 & \ion{Fe}{3} & {\it +0.29} & +0.02 & $-$0.2\tablenotemark{b} & +0.14  & 
	+0.2\tablenotemark{e} & \nodata & +0.32 & {\it $-$0.50} \\  

2057.93 & \ion{Fe}{3} & $-$0.40 & $-$0.23 & $-$0.5\tablenotemark{c} & 0.00  & 
	\nodata & $-$0.20\tablenotemark{h} & $-$0.03 & $-$0.72 \\

2058.20 & \ion{Fe}{3} & $-$0.38 & {\it $-$0.45} & $-$0.5\tablenotemark{c} & {\it $-$0.37} & 
	\nodata & $-$0.20\tablenotemark{h} &  0.00 & $-$0.90 \\

2058.55 & \ion{Fe}{3}+\ion{Cr}{3} & +0.10   & 0.00 & $-$0.1 & {\it +0.50}  & 
	\nodata & {\it +0.70} & +0.40 & {\it $-$0.65} \\

2059.67 & \ion{Fe}{3}+\ion{Mn}{3} & 0.00   & 0.00 & $-$0.1 & +0.10 & 
	 0.0 & {\it +0.70} & +0.45 & $-$0.99 \\

2063.40 & \ion{Mn}{3} & \nodata & $-$0.18 & \nodata & $-$0.01 & 
	 0.0 & $-$0.20 & +0.05 & \nodata \\

2065.26 & \ion{Fe}{3}x4+\ion{Fe}{5} & \nodata & $-$0.30 & \nodata & $-$0.10  & 
	\nodata & $-$0.40 & $-$0.10 & $-$0.85 \\

2066.40 & \ion{Mn}{3}x2+\ion{Ni}{3} & $-$0.30  & $-$0.40 & $-$0.2 & $-$0.20 & 
	\nodata & $-$0.35 & +0.05 & $-$0.90 \\

2068.25 & \ion{Fe}{3} & {\it +0.60} & +0.10 & +0.2  & {\it +0.30} & 
	{\it +0.5} & {\it +0.43} & +0.47 & $-$1.00 \\

2069.82 & \ion{Fe}{3}+\ion{Mn}{3} & +0.05   & $-$0.20 & 0.00 & $-$0.4  & 
	\nodata & $-$0.15 &  +0.20 & \nodata \\

2070.56 & \ion{Fe}{3} & {\it +0.47} & 0.00 & $-$0.3  & {\it +0.44} & 
	$-$0.1\tablenotemark{f} & {\it +0.89} & {\it +0.55} & $-$0.79 \\

2070.98 & \ion{Fe}{3} & $-$0.35   & $-$0.35  & $-$0.5  & $-$0.19   & 
	$-$0.1\tablenotemark{f} & $-$0.30 & $-$0.12 & $-$0.70 \\

2074.23 & \ion{Fe}{3} & $-$0.15    & {\it $-$0.70} & \nodata  & +0.12 & 
	\nodata & $-$0.35 & +0.14 & \nodata \\  

2076.31 & \ion{Fe}{3} & \nodata   & {\it $-$0.70} & \nodata  & $-$0.09 
	& \nodata & $-$0.40 & +0.05 & \nodata \\
\\
AVG        & & $-$0.16 & $-$0.15 & $-$0.2 & 0.02 & 0.1 & $-$0.16 & 0.16 & $-$0.88 \\ 
1 $\sigma$ & &    0.17 & 0.15    &  0.2   & 0.12 & 0.1 &    0.23 & 0.14 &    0.10 \\
\enddata
\tablecomments{
Abundances are determined from spectrum syntheses using the
atmospheric parameters in Tables~\ref{atms} and \ref{abus}.  
Dominant features only are identified here.  
Results that are $\ge$2\,$\sigma$ from the mean are {\it italicized} 
and not included in the average.
We find a mean iron-group abundance of [Fe/H]=$-$0.06 $\pm$0.15
for the main sequence B-type stars in this Table.
This is expected for solar neighborhood objects, and similar 
to the B-type star results in Cunha \& Lambert (1994) and 
Gies \& Lambert (1992). }
\tablenotetext{a,b,c,d,e,f,g,h}{The abundances noted are from a blend of 
more than one feature listed in the table, e.g., all entries marked by 
the letter `a' are from a single fit to a blend of those features.}
\tablenotetext{i}{The Ekberg (1993) wavelength does not fit the feature in
PG\,0832+676, although it does fit the features in the other sharp lined stars.}
\end{deluxetable}
\clearpage

% "deluxetable" .
\clearpage
\begin{deluxetable}{lrrrrr} 
%\footnotesize
\tablecaption{Representative Iron-group Abundance Uncertainties \label{metal-unc}}
\tablewidth{0pt}
\tablehead{
\colhead{} & \colhead{BD+56\,576} & \colhead{HD\,34078} & \colhead{HD\,36591} &  
\colhead{HD\,216916} & \colhead{PG\,0832} \\[.2ex]
\colhead{} & \colhead{$\Delta$[$M$/H]} & \colhead{$\Delta$[$M$/H]} & \colhead{$\Delta$[$M$/H]} &  
\colhead{$\Delta$[$M$/H]} & \colhead{$\Delta$[$M$/H]} 
} 
\startdata
$\Delta$\teff=$\pm$750~K          & $\mp$0.01 & $\pm$0.2 & $\pm$0.10 & $\pm$0.01 & $\mp$0.04 \nl
$\Delta$\logg=$\pm$0.1            & $\pm$0.02 & $\mp$0.1 & $\mp$0.01 & $\pm$0.03 & $\pm$0.03 \nl
$\Delta\xi=\pm$1~\kms & $\mp$0.27 & $\mp$0.1 & $\mp$0.23 & $\mp$0.18 & $\mp$0.23 \nl
$\Delta\xi_{Ma}=\pm$2~\kms & $\pm$0.09 & $\pm$0.0 & $\pm$0.11 & $\pm$0.11 & $\pm$0.14 \nl
Shift Continuum $\pm$1\%          & $\mp$0.07 & $\mp$0.1 & $\mp$0.07 & $\mp$0.08 & $\mp$0.07 \nl
\enddata
\end{deluxetable}
\clearpage

% "deluxetable" .
\clearpage
\begin{deluxetable}{lrrrrr}
%\footnotesize
\tablecaption{ Boron Abundance Uncertainties \label{boron-unc}}
\tablewidth{0pt}
\tablehead{
\colhead{} & \colhead{BD+56\,576} & \colhead{HD\,34078} & \colhead{HD\,36591} & 
\colhead{HD\,216916} & \colhead{PG\,0832}\\[.2ex]
\colhead{} & \colhead{$\Delta$log($B$/H)} & \colhead{$\Delta$log($B$/H)} & 
\colhead{$\Delta$log($B$/H)} & \colhead{$\Delta$log($B$/H)} & \colhead{$\Delta$log($B$/H)} 
} 
\startdata
$\Delta$\teff=$\pm$750~K          & $\pm$0.01 & $\pm$0.2 & $\pm$0.05 & $\pm$0.00 & $\mp$0.01 \nl
$\Delta$\logg=$\pm$0.1            & $\pm$0.03 & $\mp$0.1 & $\pm$0.03 & $\pm$0.04 & $\pm$0.05 \nl
$\Delta\xi=\pm$1~\kms & $\mp$0.03 & $\pm$0.0 & $\pm$0.00 & $\pm$0.02 & $\pm$0.00 \nl
$\Delta\xi_{Ma}=\pm$2~\kms & $\pm$0.02 & $\pm$0.0 & $\pm$0.00 & $\pm$0.04 & $\pm$0.01 \nl 
Shift Continuum $\pm$1\%          & $\mp$0.03 & $\mp$0.1 & $\mp$0.06 & $\mp$0.03 & $\mp$0.13 \nl
$^{11}$B =2\,$^{10}$B             & $-$0.02 &  0.0 & $-$0.06 & $-$0.12 & $-$0.10 \nl
$\Delta$V$_{\rm rad}=\pm$2~\kms   & $\pm$0.10 & $\pm$0.0 & $\pm$0.07 & $\pm$0.03 & $\pm$0.04 \nl
($\Delta$V$_{\rm rad}$ \ion{Mn}{3}$\lambda$2065.9)\tablenotemark{^*} & ({\it $\mp$0.10}) & ({\it $\pm$0.0}) 
	& ({\it $\mp$0.13}) & ({\it $\mp$0.26}) & ({\it $\mp$0.00}) \nl 
\enddata
\tablecomments{
$\Delta$B was determined by scaling the \ion{Mn}{3}~$\lambda$2065.9 line 
abundance {\it a priori} according to the results listed for the five 
uncertainties examined for the iron-group elements in Table~\ref{metal-unc}.
We also examine a smaller $^{11}$B/$^{10}$B ratio since this may be affected
by rotational mixing (discussed in Section~4).}
\tablenotetext{^*}{Changing the radial velocity required a completely new fit,
thus the new ``best'' \ion{Mn}{3} 2065.9\,\AA\ abundance is also listed here 
({\it in italics}).}
\end{deluxetable}
\clearpage

% "deluxetable" .
\clearpage
\begin{deluxetable}{llrrrrlrrrrl}
%\footnotesize
\tablecaption{NLTE Boron and CNO Abundances in B-type Stars \label{bcno}}
\tablewidth{0pt}
\tablehead{
\colhead{Star} & 
\colhead{-----} & \colhead{Liter} & \colhead{ature} & \colhead{-----} & \colhead{REFS} &
\colhead{-----} & \colhead{Corr} & \colhead{ected} & \colhead{-----} &
\colhead{\teff} \\[.2ex]
\colhead{} & 
\colhead{B} & \colhead{C} & \colhead{N} & \colhead{O} & \colhead{} &
\colhead{B} & \colhead{C} & \colhead{N} & \colhead{O} &
\colhead{}  
} 
\startdata
THIS PAPER:\nl
BD+56\,576 & \nodata & 7.89 & 7.35 & 8.57 & V+00 &
	2.25\tablenotemark{a} & 7.84 & 7.62 & 8.34 & 22500 \nl
HD\,16582  & $\le$1.3 & 8.14 & 8.12 & \nodata & PQ01, GL92 &
	1.16\tablenotemark{a} & 8.15 & 8.10 & \nodata & 22942 \nl
HD\,36591  & \nodata & 8.24 & 7.64 & 8.64 & GL92 &
	$\le$1.32\tablenotemark{a} & 8.25 & 7.69 & 8.67 & 26449 \nl 
---        & $\le$2.0 & 8.25 & 7.58 & 8.60 & PQ01, CL94 &
	\nodata &  8.32 & 7.64 & 8.54 & 26330 \nl
HD\,50707  & $\le$1.7 & 8.20 & 8.23 & 8.87 & PQ01, GL92 &
	$\le$1.5\tablenotemark{a} & 8.21 & 8.29 & 8.89  & 26768 \nl
HD\,205021 & $\le$1.2 & 7.96 & 7.97 & 8.77 & PQ01, GL92 &
	$\le$0.9\tablenotemark{a} & 7.98 & 8.00 & 8.81 & 25831 \nl
HD\,216916 & 2.4 & 8.15 & 7.66 & 8.64 & PQ01, GL92 &
	2.31\tablenotemark{a} & 8.17 & 7.64 & 8.61  & 23232 \nl
\nl
GHRS BORON:\nl
HD\,886    & 2.2 & 8.41 & 7.78 & \nodata & PQ01, GL92 &
	2.2 & 8.40 & 7.75 & \nodata &  21899 \nl
---        & 2.29\tablenotemark{b} & \nodata & \nodata & \nodata & P+99 &
             2.28\tablenotemark{b} & \nodata & \nodata & \nodata & 22670 \nl
HD\,3360   & $\le$1.2 & 8.43 & 8.26 & \nodata & PQ01, GL92 &
	     $\le$1.2 & 8.40 & 8.22 & \nodata & 21426 \nl
---        & 0.92\tablenotemark{b} & \nodata & \nodata & \nodata & P+99 &
             0.91\tablenotemark{b} & \nodata & \nodata & \nodata & 22180 \nl
HD\,35299  & 2.9 & 8.17 & 7.71 & 8.63 & PQ01, GL92 &
	2.9 & 8.19 & 7.70 & 8.64 & 23831 \nl
---        & 2.7 & 8.38 & 7.74 & 8.57 &  L+00, CL94 &
	2.7 & 8.40 & 7.67 & 8.37 & 24000 \nl  
---        & 2.46\tablenotemark{b} & \nodata & \nodata & \nodata & P+99 &
             2.41\tablenotemark{b} & \nodata & \nodata & \nodata & 24670 \nl
---        & 2.84\tablenotemark{c} & \nodata & \nodata & \nodata & C+97 &
             2.84\tablenotemark{c} & \nodata & \nodata & \nodata & 24000 \nl
HD\,35039  & 2.6 & 8.40 & 7.76 & \nodata & PQ01, GL92 &
	2.6 & 8.35 & 7.70 & \nodata &  20547 \nl
---        & 2.92\tablenotemark{c} & 8.50 & 7.85 & 8.60 & C+97, CL94 &
	2.92\tablenotemark{c} & 8.36 & 7.65 &  8.34 & 20550 \nl
HD\,36285  & 2.43\tablenotemark{c} & 
	8.57 &  7.95 & 8.80 & C+97, CL94 &
	2.43\tablenotemark{c} & 8.48 & 7.77 & 8.55 & 21930 \nl
---        & 1.8 & \nodata & \nodata & \nodata & PQ01 &
             1.8 & \nodata & \nodata & \nodata & 
	     {\it 22230}\tablenotemark{d} \nl
HD\,36430  & 2.53\tablenotemark{c} & 8.54 & 7.89 & 8.84 & C+97, CL94 &
        2.53\tablenotemark{c} &  8.38 & 7.67 & 8.57 & 19640 \nl 	
---        & 2.5 & \nodata & \nodata & \nodata & PQ01 &
             2.5 & \nodata & \nodata & \nodata & 19640 \nl
\nl
\tablebreak
\nl
IUE BORON:\nl
HD\,22951  & 1.8 & 8.14 & 7.69 & 8.45 & PQ01, GL92 &
	1.8 & 8.11 & 7.69 & 8.42 & 27869 \nl 
HD\,29248  & 2.5 & 8.25 & 7.75 & 8.70 & PQ01, GL92 &
	2.5 & 8.27 & 7.74 & 8.67 &  23290 \nl
HD\,30836  & \nodata & 8.44 & 7.79 & \nodata & GL92 &
	 \nodata & 8.41 & 7.75 & \nodata & 21368 \nl
---        & $\le$1.3 & \nodata & \nodata & \nodata & PQ01 & 
     $\le$1.3 & \nodata & \nodata & \nodata & {\it 20819}\tablenotemark{d} \nl
HD\,34816  & 2.3 & 8.26 & 7.66 & 8.75 & PQ01, GL92 &
	2.3 & 8.17 & 7.59 & 8.67 & 28875 \nl
HD\,35337  & 2.1 & 8.29 & 7.65 & 8.56 & PQ01, GL92 &
	2.1 & 8.31 & 7.64 & 8.55 & 23590 \nl
HD\,35468  & $\le$1.0 & 8.30 & 8.13 & \nodata & PQ01, GL92 &
	$\le$1.0 & 8.28 & 8.09 & \nodata &  21803 \nl
HD\,36351  & 2.6 &  8.35 & 7.83 &  8.76 & PQ01, CL94 &
	2.6 & 8.28 & 7.68 & 8.52 & 21950 \nl
HD\,36629  & 2.5 & 8.38 & 7.75 & 8.55 & PQ01, CL94 &
	2.5 & 8.32 & 7.61 & 8.32 & 22300 \nl
HD\,36959  & \nodata & 8.10 & 7.76 & 8.65 & GL92 &
	\nodata & 8.13 & 7.76 & 8.69 &  24517 \nl
---      & 2.5 & 8.33 & 7.76 &  8.76 & PQ01, CL94 &
	2.5 & 8.37 & 7.73 & 8.61 &  24890 \nl
HD\,36960  & \nodata & 8.27 & 7.72 & 8.88 & GL92 &
        \nodata & 8.18 & 7.65 & 8.80 & 28941 \nl
---       & 1.9 & 8.36 & 7.50 & 8.72 & PQ01, CL94 &
	1.9 & 8.39 & 7.54 & 8.71 & 28920 \nl
HD\,37209  & \nodata & 8.17 & 7.57 & 8.50 & GL92 &
	\nodata & 8.19 & 7.56 & 8.51 &  24053 \nl
---      & 2.5 & 8.27 & 7.63 &  8.83 &  PQ01, CL94 & 
	2.5 & 8.29 & 7.55 & 8.63 &  24050 \nl
HD\,37356  & 2.5 & 8.46 &  7.84 & 8.67 & PQ01, CL94 &
	2.5 & 8.41 & 7.70 & 8.44 & 22370 \nl
HD\,37481  & $\le$2.5 & 8.40 & 7.65 &  8.96 & PQ01, CL94 &
	$\le$2.5 & 8.39 & 7.55 & 8.75 &  23300 \nl
HD\,37744  & 2.5 & 8.34 & 7.85 & 8.63 & PQ01, CL94 &
	2.5 & 8.37 & 7.80 & 8.46 &  24480 \nl
HD\,41753  & 2.5 & 8.58 & 8.15 & \nodata & PQ01, GL92 &
        2.5 & 8.53 & 8.10 & \nodata &  17272 \nl
HD\,44743  & 2.8 & 8.21 & 7.71 & 8.75 & PQ01, GL92 &
	2.8 & 8.23 & 7.73 & 8.78 &  25725 \nl
HD\,46328  & $\le$1.4 & 8.04 & 7.89 & 8.69 & PQ01, GL92 &
	$\le$1.4 & 8.05 & 7.95 & 8.71 & 26778 \nl
HD\,52089  & $\le$1.6 & 8.25 & 8.12 & 8.50 & PQ01, GL92 &
	$\le$1.6 & 8.27 & 8.04 & 8.30 & 23908 \nl 
HD\,184171 & 1.8 & 8.32 & 7.91 & \nodata & PQ01, GL92 &
	1.8 & 8.28 & 7.86 & \nodata & 16557 \nl
HD\,214993 & 2.3 & 8.25 & 7.80 & 8.78 & PQ01, GL92 &
	2.3 & 8.28 & 7.82 & 8.83 & 24759 \nl
\tablebreak
\enddata
\tablecomments{This table summarizes the NLTE B, C, N, and O data 
for B-type main sequence stars, excluding stars with 
18,000 $\le$ \teff\ $\le$ 29,000\,K (see text).  
The left side of the table quotes NLTE abundances as published 
in the literature.   The right side of the table corrects the NLTE 
CNO abundances by (i) including adjustments to correct for the use 
of Gold, instead of Kurucz, model atmospheres, and (ii) reducing 
the GL92 temperatures by 3.4\%, and adjusting their abundances 
according to the offsets listed in their Table~9.   This reduction 
in the GL92 temperatures has a negligible effect on the \ion{B}{3} 
abundances (e.g., see Table~\ref{boron-unc}).
There are some significant differences between the GL92 and CL94 
abundances for the five stars in common.  The differences in oxygen 
appear to be due to differences in $\xi$, whereas those in C appear 
when two strong \ion{C}{3} lines are included by GL92.   N is in
good agreement throughout, with the only significant difference 
of $\sim$0.1~dex occuring for the hot star HD\,36960. 
}
\tablenotetext{a}{Boron abundances from this paper.}
\tablenotetext{b}{Proffitt \etal (1999) LTE \ion{B}{3} abundances
have been NLTE corrected (corrections are from this paper, see 
Table~\ref{boron-unc}) for this table.   On the right side of the table, 
their abundances are also corrected for the differences between their 
temperatures and the reduced GL92 temperatures listed (small effect).}
\tablenotetext{c}{Boron from \ion{B}{2} 1362.5\,\AA, which is severely
blended with metal lines.}
\tablenotetext{d}{PQ01 temperatures are the same as CL94 or they have
adopted GL92$-$820~K.  This offset is very similar to our reduction of 
3.4\%.   We note when the PQ01 and GL92 temperatures differ by more than 250~K.} 
\end{deluxetable}
\clearpage

%% Input the Figures, FIGURES, figures

\clearpage
\clearpage

\clearpage
\clearpage

\clearpage
\clearpage

\clearpage
\clearpage

\clearpage
\clearpage

\clearpage
\clearpage

\clearpage
\clearpage

\clearpage
\clearpage

\clearpage
\clearpage

\clearpage
\clearpage

\clearpage
\clearpage

\clearpage
\clearpage

\clearpage
\clearpage

\clearpage
\clearpage

\end{document}